\documentclass{aa}  

\usepackage{graphicx}
\usepackage{txfonts}
\usepackage{color}
\usepackage{comment}

\newcommand{\Msun}{M_{\odot}}

\newcommand{\Omstar}{\Omega_{*}}
\newcommand{\Omsun}{\Omega_{\odot}}
\newcommand{\Mjup}{M_{\mathrm{Jup}}}
\newcommand{\omtide}{\omega_{\mathrm{tide}}}
\newcommand{\Gwind}{\Gamma_{\mathrm{wind}}}
\newcommand{\Gint}{\Gamma_{\mathrm{int}}}
\newcommand{\Gtide}{\Gamma_{\mathrm{tide}}}
\newcommand{\Teff}{T_{\mathrm{eff}}}
\newcommand{\Qeq}{Q'_{\mathrm{eq}}}
\newcommand{\Qdyn}{Q'_{\mathrm{dyn}}}
\newcommand{\Pini}{P_{*,\mathrm{ini}}}
\newcommand{\aini}{a_{\mathrm{ini}}}
\newcommand{\Porb}{P_{\mathrm{orb}}}
\newcommand{\tint}{\tau_{\mathrm{int}}}
\newcommand{\Ro}{R_{\mathrm{o}}}
\newcommand{\Rof}{R_{\mathrm{of}}}
\newcommand{\Rosun}{R_{\mathrm{o},\odot}}
\newcommand{\Rosat}{R_{\mathrm{o,sat}}}

\begin{document}

   \title{Evolution of star-planet systems\\under magnetic braking and tidal interaction}
   \subtitle{}

   \author{M.~Benbakoura\inst{\ref{inst:aim}, \ref{inst:aim2}, \ref{inst:syrte}} \and V.~Réville\inst{\ref{inst:ucla}, \ref{inst:aim}, \ref{inst:aim2}} \and A.~S.~ Brun\inst{\ref{inst:aim}, \ref{inst:aim2}} \and C.~Le~Poncin-Lafitte\inst{\ref{inst:syrte}}\and S.~Mathis\inst{\ref{inst:aim}, \ref{inst:aim2}}}

   \institute{IRFU, CEA, Université Paris-Saclay, F-91191 Gif-sur-Yvette Cedex, France\\
   \email{mansour.benbakoura@cea.fr}\label{inst:aim}
         \and
         Université Paris Diderot, AIM, Sorbonne Paris Cité, CEA, CNRS, F-91191 Gif-sur-Yvette Cedex, France\label{inst:aim2}
         \and SYRTE, Observatoire de Paris, Université PSL, CNRS, Sorbonne Universités, LNE, 61 Avenue de l’Observatoire, 75014 Paris, France \label{inst:syrte}
         \and UCLA Earth, Planetary and Spaces Sciences, 595 Charles Young Drive East, Los Angeles CA 90035, United States\label{inst:ucla}
         }

   \date{Received XXXX; accepted YYYY}

 
  \abstract
   {With the discovery over the last two decades of a large diversity of exoplanetary systems, it is now of prime importance to characterize star-planet interactions and how such systems evolve.}
   {We address this question by studying systems formed by a solar-like star and a close-in planet. We focus on the stellar wind spinning down the star along its main sequence phase and tidal interaction causing orbital evolution of the systems. Despite recent significant advances in these fields, all current models use parametric descriptions to study at least one of these effects. Our objective is to introduce simultaneously ab-initio prescriptions of the tidal and braking torques, so as to improve our understanding of the underlying physics.}
   {We develop a 1D numerical model of coplanar circular star-planet systems taking into account stellar structural changes, wind braking and tidal interaction and implement it in a code called ESPEM. We follow the secular evolution of the stellar rotation assuming a bi-layer internal structure, and of the semi-major axis of the orbit. After comparing our predictions to recent observations and models, we perform tests to emphasize the contribution of ab-initio prescriptions. Finally, we isolate four significant characteristics of star-planet systems: stellar mass, initial stellar rotation period, planetary mass and initial semi-major axis; and browse the parameter space to investigate the influence of each of them on the fate of the system.}
   {Our secular model of stellar wind braking reproduces well the recent observations of stellar rotation in open clusters. Our results show that a planet can affect the rotation of its host star and that the resulting spin-up or spin-down depends on the orbital semi-major axis and on the joint influence of magnetic and tidal effects. The ab-initio prescription for tidal dissipation that we used predicts fast outward migration of massive planet orbiting fast-rotating young stars. Finally, we provide the reader with a criterion based on the system's characteristics that allows us to assess whether or not the planet will undergo orbital decay due to tidal interaction.}
   {}


	\titlerunning{Evolution of star-planet systems}
	\maketitle
%

\section{Introduction}

The planetary systems discovered during the last two decades show a wide and unexpected diversity. Indeed, the detection of 51 Pegasi b \citep{MayorQueloz1995}, a Jupiter-like planet orbiting around its star in less than five days, questioned the theories of planetary systems formation which were based on observations of the Solar System. Recently, the discovery of Proxima b \citep{AngladaEscudeetal2016} and the planetary system of TRAPPIST-1 \citep{Gillonetal2017} paved the way for research of habitable planets around low-mass stars. Understanding how such systems form and evolve is one of the most challenging questions in astrophysics.

A large proportion of systems where one planet or more is orbiting closer to its host star than Mercury to the Sun have been observed. Tidal interactions play a key role in the orbital configuration of these very compact systems since it is likely to circularize orbits, align spins and synchronize periods \citep{Zahn1977,MathisRemus2013,Ogilvie2014}. It consists in an exchange of angular momentum between the orbit and the spins of the celestial bodies. This exchange is the consequence of the dissipation of tidal flows. Their kinetic energy is converted into heat through tidal dissipation. Since the planet is synchronized within a timescale of a few thounsands of years, the stellar tide drives the secular orbital evolution \citep{Guillotetal1996,Rasioetal1996,Leconteetal2010}. In this work, we neglected the impact of the dissipation in the radiative zone. In stellar convection zones, there are two kinds of tides and both are dissipated by the turbulent friction applied by convective eddies. On the one hand, the equilibrium tide is the large-scale velocity field associated with tidal deformation, the so-called tidal bulge. This non-wavelike entity corresponds to the hydrostatic adjustment of the star to the gravitational perturbation \citep{Zahn1966b,Remusetal2012}. The friction applied by convective motions delays the response of the star to the perturbation \citep[e. g.][]{Zahn1989,OgilvieLesur2012,Mathisetal2016}. This results in a lag angle between the axes of the tidal bulge and the line of centers. This angle increases with dissipation magnitude. \citet{Hansen2012} calibrated its value for several stellar masses by constraining the dissipation using observations of planetary systems. Since lower-mass stars have deeper convective envelopes, they dissipate more energy than higher-mass stars. On the other hand, in rotating bodies such as stars, at low tidal frequencies, the Coriolis acceleration acting on this equlibrium tide excites inertial modes \citep{OgilvieLin2007}. Their ensemble, the dynamical tide, constitutes the wavelike part of the tidal response. Its dissipation strongly depends on internal structure since it arises from their reflection on the radiative, stably stratified core \citep{Ogilvie2013,Mathis2015}. It may also vary over several orders of magnitude with rotation since inertial waves are restored by the Coriolis force. At low frequencies, dissipation of the dynamical tide is several orders of magnitude higher than the dissipation of the equilibrium tide \citep{OgilvieLin2007}.

Orbital evolution occurs simultaneously with variations of stellar structure and rotation \citep{GalletBouvier2015,Amardetal2016,Bolmontetal2012,BolmontMathis2016}. In the case of solar-like stars, the latter is slowed down over most of the main sequence by magnetic braking. This phenomenon occurs because of the wind carrying angular momentum away from the star \citep{Schatzman1962,WeberDavis1967}. Models of secular evolution of stellar rotation generally consider that it undergoes three main phases. First, before the disk dissipates, stellar rotation remains constant. The physical processes that balance accretion and contraction have long been thought to have a magnetic origin. \citet{MattPudritz2005} and \citet{Mattetal2012a} investigated the braking caused by accretion-powered stellar winds while \citet{ZanniFerreira2013} studied the effect of magnetic ejections on stellar spin. Recently, \citet{BouvierCebron2015} explored the possibility that tidal and magnetic interaction with a close-in planet embedded in the disk could compete with accretion and contraction. After the disk dissipates, the star spins up due to its contraction during the pre-main sequence \citep[see][and references therein]{Amardetal2016}. Finally, once on the main sequence (MS), the star spins down under magnetic braking. Observations show that rotational velocities of young stars range from one to a hundred times the solar velocity whereas evolved stars tend to converge to the solar rate on the Skumanich sequence \citep{Skumanich1972}. \citet{GalletBouvier2015} investigated the mass-dependence of rotational evolution and showed that the braking torque and the core-envelope coupling timescale strongly depend on stellar mass. \citet{Mattetal2015} use the Rossby number to disentangle solar-like stars populations in two groups, fast, saturated rotators and slow, unsaturated, rotators. They showed that the spin-down timescale was decreasing with stellar mass in the former group and increasing in the latter, in agreement with the observations of \citet{Barnes2010}.

In star-planet systems, the interactions between the central body and its companion result in intricate phenomena that involve the star's structure, rotation and magnetism and the planet's orbital parameters. For example, the spin-down of a solar-like star over the main sequence increases its co-rotation radius and this may occur until the latter becomes larger than the orbital semi-major axis, causing a change of sign of the tidal torque. Moreover, the planet spiralling inward may spin up its host star, thus impacting its magnetism through dynamo processes \citep[e. g.][]{Brunetal2004,Brunetal2015}. Recent numerical and theoretical works have allowed significant advances in our comprehension of these mechanisms. \citet{Dobbs-Dixonetal2004} studied the combined effects of tidal dissipation and magnetic braking to explain the observed distributions of orbital eccentricities. \citet{BarkerOgilvie2009} investigated the influence of these effects on spin alignment. \citet{Ferraz-Melloetal2015} used the creep tide theory from \citet{Ferraz-Mello2013} and a semi-analytical wind model \citep[see][and references therein]{Bouvier2013} to compute the past evolution of observed star-planet systems. \citet{DamianiLanza2015} assumed a constant tidal efficiency and a Skumanich-type wind braking law, in which the torque is proportional to the stellar rotation cubed, and demonstrated that a pseudo-stable equilibrium state can exist for star-planet systems, in which co-rotation is not achieved and the ratio of the orbital mean motion divided by the stellar rotation rate is determined by the angular momentum loss rate due to magnetic braking. \citet{ZhangPenev2014} implemented a star-planet system secular evolution code based on the two-layer rotational model of \citet{MacGregorBrenner1991} and the constant quality factor framework from \citet{Goldreich1963} and performed a statistical analysis on their numerical simulations to constrain the tidal theory. \citet{BolmontMathis2016} adapted the frequency-averaged results of \citet{Mathis2015} to the time-lag framework \citep{Mignard1979} and studied its impact on secular evolution of star-planet systems when coupled with a one-layer rotational model. Despite the insights they have given for the comprehension of planetary systems, all the precited works relied on a parametrized description of tidal dissipation or wind braking.

In this first study, we consider systems constituted by a solar-like star and a planet. We characterize the secular evolution of stellar rotation and orbital parameters under magnetic braking and tidal dissipation. Following \citet{ZhangPenev2014} and \citet{BolmontMathis2016}, we aim at developing a tidal model that takes into account the dissipation of both equilibrium and dynamical tides in the convective envelope and their dependence on stellar structure and rotation. We use a two-layer model of stellar interior \citep{MacGregorBrenner1991,MacGregor1991} to study the evolution of rotation. In section~\ref{sec:model}, we present the hypotheses of our study and detail the interactions that take seat in the considered system. In section~\ref{sec:rotation}, we explain how we computed the torque of \citet{Revilleetal2015a} as a function of structural and dynamical parameters of the star and compare our results with the model of \citet{Mattetal2015}. To quantify the tidal torque, we used the empirically calibrated results of \citet{Hansen2012} for the equilibrium tide and the theoretical prediction of \citet{Mathis2015} for the dynamical tide, as discussed in section~\ref{sec:tides}. In section~\ref{sec:results}, we use ESPEM (French acronym for Evolution of Planetary Systems and Magnetism) to compare our model to those of the literature and explore the influence of the fundamental characteristics of systems on their fate, that is to determine the survival of the exoplanet and its influence on stellar rotation. Finally, in section~\ref{sec:conclusion}, we summarize our results and detail the perspectives opened by this work.

\section{Star-planet interaction model}
\label{sec:model}

\begin{figure}
\centering
\includegraphics[width=\hsize]{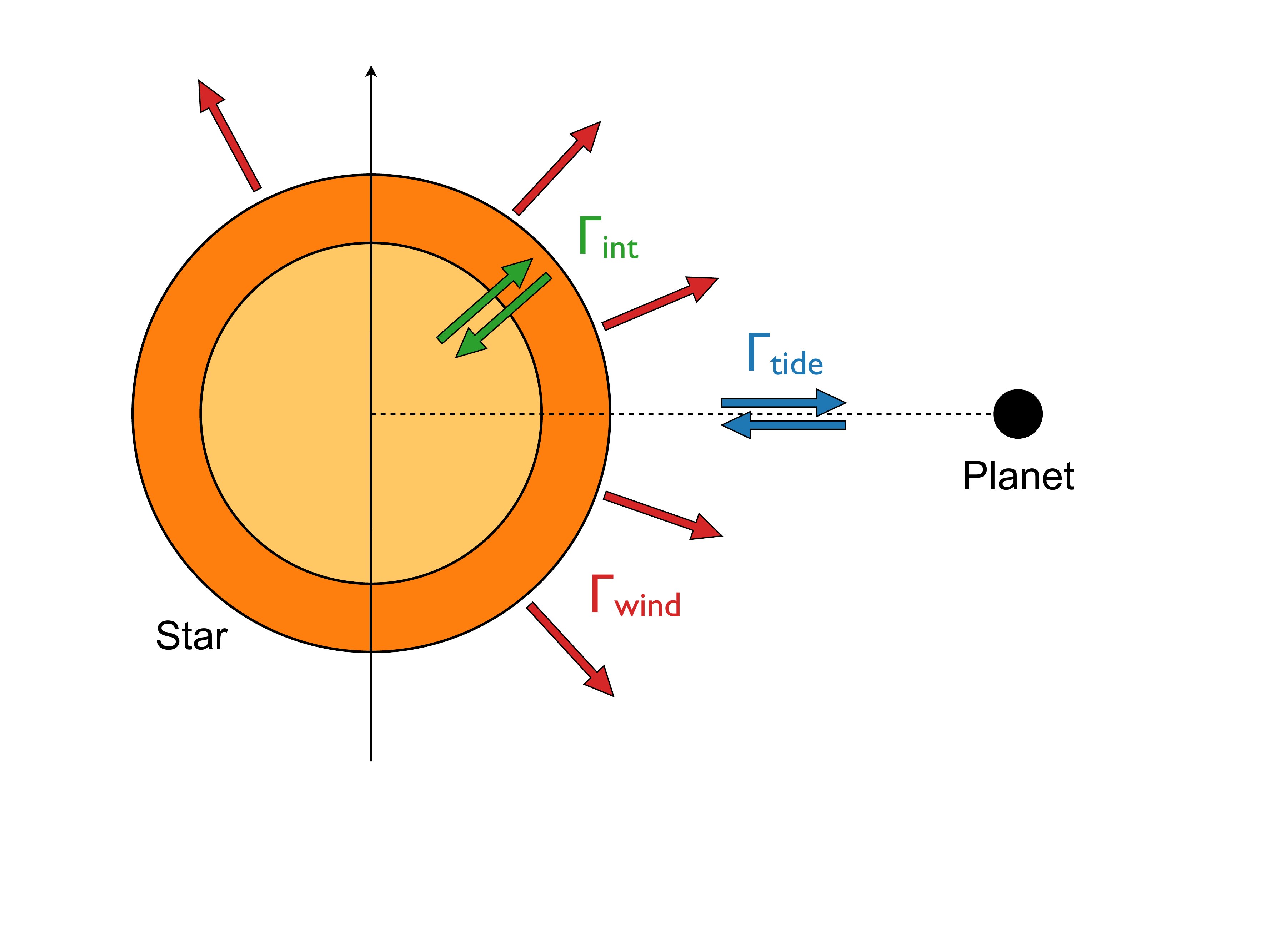}
\caption{Schematic view of the system and its interacting entities. The yellow disk represents the radiative zone and the orange shell, the convective envelope. The red arrows represent the loss of angular momentum due to stellar wind. The green arrows correspond to the exchange of angular momentum between the core and the envelope and the blue arrows to the tidal exchange between the envelope and the planetary orbit.}
\label{fig:star_planet_scheme}
\end{figure}

We consider systems formed by a planet orbiting a solar-like star. We used the two-layer star model of \citet{MacGregorBrenner1991} to study stellar rotation. In this framework, both the radiative core and the convective envelope rotate uniformly. They exchange angular momentum and only the envelope is directly braked by the wind. Thus, the core is also braked but this depends on the choice of a coupling time scale, since we do not explicitly here solve for the physical mechanisms (magnetic field, waves, turbulence) that are likely to do so \citep[see for instance][and references therein]{RudigerKitchatinov1997,Spadaetal2010,Brunetal2011,Strugareketal2011b,TalonCharbonnel2005,SpiegelZahn1992}. The basic idea of this model is that, in absence of any perturbation, the core and the envelope evolve towards synchronization of their spins. The amount of angular momentum $\Delta L$ that the core and the envelope should exchange to achieve this final state, positive if the core gives it to the envelope, negative in the other case, is:
\begin{equation}
\Delta L = \dfrac{I_c I_r}{I_c + I_r} (\Omega_r - \Omega_c), \label{eq:couplingMGB91}
\end{equation}
where $I_c$ and $\Omega_c$ (respectively $I_r$ and $\Omega_r$) are the moment of inertia and rotational velocity of the convective (respectively radiative) zone. The timescale of angular momentum redistribution, $\tau_{\mathrm{int}}$, is a free parameter of the model.

Another process occurring within the star causes angular momentum exchange between the core and the envelope. During the PMS, while the radiative zone is growing in size and mass, shells of matter rotating at the same velocity as the envelope are deposited on the core, which results in an angular momentum transfer from the envelope to the core. Thus, the internal torque $\Gint$ corresponding to the angular momentum exchange per time unit given by the radiative zone to the convective zone is given by:
\begin{equation}
\Gamma_{\mathrm{int}} = \dfrac{\Delta L}{\tau_{\mathrm{int}}} - \frac{2}{3} R_r^2 \Omega_c \dfrac{\mathrm{d} M_r}{\mathrm{d} t}, \label{eq:structureMG91}
\end{equation}
where $M_r$ and $R_r$ are the mass and radius at the base of the envelope, respectively. The first term of the sum is due to simple parametrized angular momentum redistribution while the second one is caused by stellar evolution \citep{MacGregor1991}.

We used the STAREVOL evolutionary tracks \citep{Amardetal2016} to get the one-dimensional internal structure of the star at each ESPEM time step. This model computes the stellar mass and radius, the mass and radius at the base of the envelope and the moments of inertia of the radiative and convective zones and their variations along the evolution. We studied stars of mass ranging from $0.5 M_{\odot}$ to $1.2 \ M_{\odot}$.

The orbit is assumed to be circular and coplanar and we focus on variations of the semi-major axis. Moreover, we consider that the planet's rotation is synchronized. Indeed, \citet{Guillotetal1996} showed that the synchronization of a hot Jupiter occurs within a typical timescale of $10^6$ yr, which is negligible compared to the lifetime of a star-planet system. Thus, the planet can be considered as synchronized from the beginning of the simulation. Consequently, in our model, only tidal dissipation within the star impacts orbital evolution. As a simplification for this first study, we assume that this process takes place in the envelope, which involves that the tidal torque is applied on the convective zone only, and not on the core. In our future works, the dynamical tide in the radiative zone, constituted of gravity waves dissipated by thermal diffusion and breaking will have to be implemented in our code \citep{Zahn1975,GoldreichNicholson1989,Terquemetal1998,OgilvieLin2007,BarkerOgilvie2010}. In this first step, we focus on the convective zone, which allows consistent comparison with \citet{ZhangPenev2014} and \citet{BolmontMathis2016}.

The core is interacting with the envelope through internal coupling as detailed in equations \eqref{eq:couplingMGB91} and \eqref{eq:structureMG91}. The orbit is exchanging angular momentum with the envelope. The latter acts as an interface between the planet and the core and loses angular momentum through magnetic braking. Consequently, the semi-major axis of the orbit $a$, the angular momentum of the convective zone $L_c$ and of the radiative zone $L_r$ are determined by the following equations:
\begin{align}
\dfrac{\mathrm{d} a}{\mathrm{d} t} &= - a^{1/2} \dfrac{2}{m_{\mathrm{p}} M_*} \sqrt{\dfrac{m_{\mathrm{p}} + M_*}{G}} \Gamma_{\mathrm{tide}} \label{eq:evol_semi-major} \\
\dfrac{\mathrm{d} L_c}{\mathrm{d} t} &= \Gamma_{\mathrm{int}} + \Gamma_{\mathrm{tide}} - \Gamma_{\mathrm{wind}}\\
\dfrac{\mathrm{d} L_r}{\mathrm{d} t} &= - \Gamma_{\mathrm{int}},
\end{align}
where $G$ is the gravitational constant, $m_{\mathrm{p}}$ the planetary mass, $M_*$ the stellar mass, $\Gwind$ and $\Gtide$ the wind braking and the tidal torques. The latter two correspond to the rates of changes of angular momentum associated with wind braking and tidal dissipation. Their calculation is presented in the following sections. We used the ODEX solver \citep{Haireretal2000} implementing the Bulirsch-Stoer algorithm to solve these equations and compute the secular evolution of stellar rotation and planetary orbit. In the following, we detail how the braking and tidal torque were computed. The different entities and their interactions are shown in a simplified view in Fig.~\ref{fig:star_planet_scheme}.

When planets come too close to their host star, they are destroyed either by tidal forces or after having plunged into the stellar atmosphere. To model this phenomenon in ESPEM, we followed \citet{ZhangPenev2014} who used the ratio of the planet's density divided by that of the star $\rho_{\mathrm{p}} / \rho_*$ to determine which scenario occurs. Their approach was based on the observational results of \citet{Metzgeretal2012}, who showed that three scenarios were likely to take place when a planet inspirals towards its host star. In the first case, $\rho_{\mathrm{p}} / \rho_* > 5$ and the planet spirals until it plunges in the stellar atmosphere. In the second case, $1 < \rho_{\mathrm{p}} / \rho_* <5$ and the planet overflows its Roche lobe before reaching the stellar surface. This results in an unstable mass transfer from the planet to the star and the former is torn apart within a timescale of several hours. In the third case, $\rho_{\mathrm{p}} / \rho_* < 1$, the planet also overflows its Roche lobe before getting in contact with its star but the mass transfer is stable. Consequently, the planet is disrupted over a much longer timescale, typically several thousand years. Even if these three cases lead to various timescales, planet destruction lasts briefly compared to stellar lifetime. This is why we consider that it occurs instantaneously. If $\rho_{\mathrm{p}} / \rho_* > 5$, the planet is removed from the simulation if the semi-major axis becomes smaller than the stellar radius. In the other two cases, the planet is destroyed if it starts orbiting the star below the Roche limit, which we calculated with the formula of \citet{ZhangPenev2014}:
\begin{equation}
r_{\mathrm{Roche}} = 2.44 \ r_{\mathrm{p}} \ \left(\frac{M_*}{m_\mathrm{p}}\right)^{1/3},
\end{equation}
where $r_{\mathrm{p}}$ is the planetary radius, $M_*$ the stellar mass and $m_{\mathrm{p}}$ the planetary mass. We adopted the assumption that the planets have the same mean density as Jupiter, that is, 1.33 g.cm$^{-3}$. This is a reasonable choice for planets more massive than $3 \times 10^{-2} \Mjup$ that are known to be gaseous \citep{Baraffeetal2014}. We only considered planets with a mass above this value, which corresponds to the upper mass limit of super Earths ($\sim 10\ M_{\oplus}$). We made this simplifying choice because the planets which raise the most significant tides within their host star are the most massive ones.

\section{Stellar rotation}
\label{sec:rotation}

In this section, we compute $\Gwind$. Magnetic braking occurs because of the stellar wind carrying angular momentum away \citep{Schatzman1962}. \citet{Parker1958} showed that the particles of the wind were accelerated along their way through the corona. The Alfvén radius $r_A$ is the distance at which their speed reaches the Alfvén speed. In a simplified one-dimensional model, \citet{WeberDavis1967} showed that the resulting angular momentum loss was equal to the product of the stellar convective zone rotation $\Omega_{c}$, the mass loss $\dot{M}$ and the Alfvén radius squared. After integration on a sphere, the equality becomes a proportionality relation:
\begin{equation}
\Gamma_{\mathrm{wind}} \propto \Omega_c \dot{M} r_A^2.
\label{eq:wind_torque_weberdavis}
\end{equation}

In this product, the only easily measurable factor is the stellar rotation. This can be done either by spectroscopy \citep{ReinersSchmitt2003} or by photometry \citep{McQuillanAigrainMazeh2013,Garciaetal2014rotation}. Such measurements of the other factors are however more difficult. This is why we used the model of \citet{Revilleetal2015a} to express them as a function of the star's structure and dynamics. Note however that other options have been used in the literature \citep[e. g.,][and references therein]{Brown2014,GalletBouvier2015,Johnstoneetal2015AA577I,Johnstoneetal2015AA577II,vanSadersetal2016,SadeghiArdestanietal2017}.

\citeauthor{Revilleetal2015a} showed that the Alfvén radius was only dependent on the magnetic flux through the open field lines, $\Phi_{\mathrm{open}}$, for any given topology:
\begin{equation}
\Gwind = \dot{M} \Omega_c R_*^2 K_3^2 \left[\frac{\Phi_{\mathrm{open}}^2 / R_*^2}{\dot{M} v_{\mathrm{esc}} \sqrt{1 + \displaystyle{\left(\frac{f}{K_4}\right)}^2}}\right]^{2m},
\label{eq:wind_torque_revilleetal}
\end{equation}
where $R_*$ is the stellar radius, $v_{\mathrm{esc}} = \sqrt{2 G M_* / R_*}$ is the escape velocity, $f = \Omega_c / \sqrt{G M_* / R_*^3}$ is the ratio of rotational velocity of the envelope over the brake-up rotation rate and the constants $K_3$, $K_4$, and $m$ were set by the authors using numerical simulations \citep{Revilleetal2015a}. There are two kinds of magnetic field lines, the closed ones, also known as magnetic loops, and the open ones. The latter coincide with the wind streamlines in the interplanetary medium. Thus, the open flux measures the magnetic flux in the wind. It depends on topology, which is the repartition of magnetic energy in the different spherical harmonics. Higher order topology magnetic fields decrease more steeply with distance from the star than lower order ones: a multipolar field of degree $l$ decreases as $1 / r^{l+2}$ where $r$ is the distance to the star. This results in smaller magnetic fluxes at equal distances. Therefore, the open flux decreases with the order of the magnetic multipole considered.

The formula of Eq.~\eqref{eq:wind_torque_revilleetal} has been obtained with 2D numerical simulations. \citet{Revilleetal2016} verified this result and revised the list of dimensionless constants ($K_3$, $K_4$, $m$) with 3D simulations constrained by realistic magnetic mappings obtained by spectropolarimetric measurements and concluded that $K_3 = 0.55$, $K_4 = 0.06$ and $m = 0.3$.

Formulating the torque of the wind as a function of the open magnetic flux allows us to write a topology-independent law. This idea was also used in the observational study of \citet{Seeetal2017}, who estimated the open flux of 66 solar-like stars from Zeeman-Doppler magnetograms. After \citet{Revilleetal2015a,Revilleetal2015b} and \citet{Garraffoetal2015}, \citet{FinleyMatt2017,FinleyMatt2018ApJ857} used the open flux to study the wind of stars with complex magnetic topologies involving interactions between the dipolar, quadrupolar, and octupolar components. In the following subsections, we describe the steps of our method to estimate the wind-braking torque from the stellar structure and rotation.

\subsection{Mass loss rate and open magnetic flux}

\begin{figure}
\centering
\includegraphics[width=\hsize]{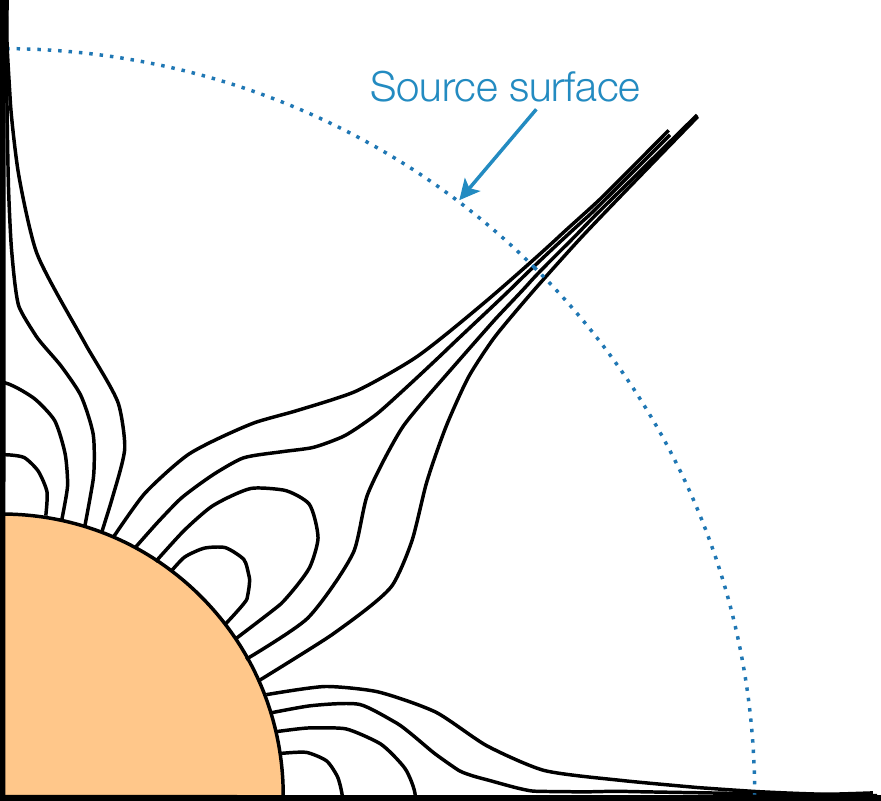}
\caption{Schematic view of a solar-type star's corona. The solid black circle delimits the photosphere. Magnetic field lines are represented as curved black lines. The source surface is shown as a dotted blue circle. All the coronal loops are contained below this surface.}
\label{fig:magnetic_scheme}
\end{figure}

Computing the torque of the wind requires to first calculate the mass-loss rate and the open magnetic flux. Since these quantities are determined by the properties of the open magnetic field lines, a consistent model of the corona is needed to infer their global properties. To that end, we used the starAML routine developed by \citet{Revilleetal2015b}, who presented a method to determine the maximal size of the main coronal streamer for a given star with known mass, radius, effective temperature, surface magnetic field, base coronal density and temperature (see Appendix~\ref{sec:descriptionCode} for more details). The result of this calculation corresponds to the radius of the surface beyond which all field lines are opened by the wind, the so-called source surface \citep{Schattenetal1969,AltschulerNewkirk1969,SchrijverDeRosa2003}. Figure~\ref{fig:magnetic_scheme} illustrates the coronal structure of a solar-type star with its magnetic structures.

The method consists in extrapolating the surface magnetic field to compute the source surface radius from the equilibrium between the ram and thermal pressures of the gas and the magnetic pressure. The mass-loss rate and open magnetic flux are then inferred from this calculation. First, the radial profiles of the temperature and density of the wind are computed assuming a polytropic equation of state of index $\gamma = 1.05$. These profiles allow us to compute the velocity field in the corona. The mass-loss rate is deduced from these computations as $\dot{M} = 4\ \pi r^2 v^2 \rho^2$, where $r$, $v$, and $\rho$ correspond to the distance to the star, velocity, and mass density at the outer boundary of our calculation grid, respectively.

Then, the magnetic field is extrapolated from the surface to the rest of the corona. From these quantities, the thermal and ram pressures of the wind can be computed in the whole corona as well as the magnetic pressure. Close to the stellar surface, the magnetic pressure dominates the other two and is sufficient to hold the magnetic field lines closed. Far from the star, the pressure difference changes sign and the ram and thermal pressures of the wind open the magnetic field lines. The surface at which this inversion occurs is defined as the source surface. If we approximate this surface by a sphere, it is possible to define its radius, $r_{ss}$. We computed it from the radial profiles of the thermal, ram, and magnetic pressures and deduced the open magnetic flux, which is equal to the flux of the magnetic field through this surface (see Fig.~\ref{fig:magnetic_scheme}). This method requires assumptions on the density $n_c$ and temperature $T_c$ at the base of the corona. These boundary conditions are detailed in the next subsection.

\subsection{Assumptions at the base of the corona}

For the solar values, we followed \citet{Revilleetal2016} and used $n_{\odot} = 10^8\ \mathrm{cm}^{-3}$ and $T_{\odot} = 1.5 \times 10^{6}\ \mathrm{K}$, which reproduce well the speed of the solar wind measured on Earth. For other stars, we followed \citet{HolzwarthJardine2007} and \citet{Revilleetal2016}, who extended these hypotheses assuming that the temperature $T_c$ and density $n_c$ at the base of the corona depended mostly on the surface rotation rate:
\begin{align}
n_c &= n_{\odot} \left(\frac{\Omega_c}{\Omega_{\odot}}\right)^{0.6}, \label{eq:corDensity} \\
T_c &= T_{\odot} \left(\frac{\Omega_c}{\Omega_{\odot}}\right)^{0.1}. \label{eq:corTemperature}
\end{align}
A more recent model on mass-loss rates of cool main-sequence stars \citep{CranmerSaar2011} predicts a slightly steeper dependence of $n_c$ on $\Omega$. However, the scatter in their results still allows for a dependence like our Eq.~\eqref{eq:corDensity}. Please note that alternative laws for the variations of coronal temperature  with respect to global stellar parameters have been recently proposed \citep[e. g.,][]{JohnstoneGudel2015,Woodetal2018}. These formulations are discussed in section~\ref{sec:conclusion}.

At fast rotations, we introduced saturation by considering that $n_c$ and $T_c$ remained constant. The following subsection describes how this saturation is determined.

\subsection{Saturation}

To determine the saturation rate, we followed the approach of \citet{Mattetal2015}, who used the stellar Rossby number to define it:
\begin{equation}
\Ro = (\Omega_c \tau_{\mathrm{conv}})^{-1}, \label{eq:stellarRossby}
\end{equation}
where $\tau_{\mathrm{conv}}$ is the convective turnover timescale, which we computed with the formula of \citet{CranmerSaar2011} that is mostly valid on the main sequence, using the value of the effective temperature at the ZAMS:
\begin{equation}
\tau_{\mathrm{conv}} = 314.24 \exp\left[- \left(\frac{\Teff^{\mathrm{ZAMS}}}{1952.5\ \mathrm{K}}\right) - \left(\frac{\Teff^{\mathrm{ZAMS}}}{6250\ \mathrm{K}}\right)^{18}\right] + 0.002. \label{eq:tauConv}
\end{equation}
We kept the effective temperature at the ZAMS in this expression because it allowed us to simplify the calculations with starAML. This is a reasonable assumption since the variations of $\Teff$ are negligible during the main-sequence phase (< 5\%). This point is further discussed at the end of section~\ref{sec:conclusion}.

The saturation value of the Rossby number was set to $\Rosat = 0.1 \ \Rosun$. The convective turnover timescale is indirectly linked to stellar mass. Indeed, among solar-type stars, lower-mass stars have thicker envelope and slower flows. Therefore, their convective cells travel on longer timescales, which is why the turnover time is a decreasing function of stellar mass. Consequently, the Rossby number contains information about rotation and mass at the same time. It is also a good indicator of magnetic activity, as shown by \citet{Noyesetal1984} who emphasized the correlation between chromospheric activity and Rossby number, and \citet{Pizzolatoetal2003} and \citet{Wrightetal2011} who found evidence of saturation of X-ray emissions at $\Ro = 0.1\ \Rosun$.

\subsection{Global surface magnetic field}

Solar-type magnetic fields are believed to be generated by a dynamo in the envelope \citep{Brunetal2004,Brunetal2015}. Therefore, their strength depends on the depth of the convective zone, which decreases with stellar mass, and the rotation of the star \citep{Noyesetal1984}. This leads us to consider the following law for the magnetic energy at the surface:
\begin{align}
B & \propto M_*^{-3.5} \ \Omega_c^{0.5} & \mbox{(saturated)} \label{eq:MagFieldSat}\\
B & \propto M_*^{-3.5} \ \Omega_c^{2} & \mbox{(unsaturated)}. \label{eq:MagFieldUnsat}
\end{align}
The solar value was set to $B_{\odot} = 3 \ \mathrm{G}$, which is in good agreement with observations \citep{Vidottoetal2014}.

We assumed a dipolar magnetic topology to compute the braking torque. The Alfvén radius from our calculations was 5.25 $R_*$ for a solar-mass star rotating at the solar rate. However, this value was not sufficient to brake the $1 M_{\odot}$ star efficiently and reproduce the convergence of rotational velocities at solar age. This could be related to the 'Open Flux Problem', which reveals that magnetohydrodynamics wind models consistent with surface observations systematically underestimate the interplanetary magnetic flux \citep[see][]{Linkeretal2017}. To fix this lack of braking, we artificially multiplied the Alfvén radius by a factor 3.6. This correction allows us to conciliate our theory with observations of open clusters.

\subsection{Braking torque}

\begin{figure}
\centering
\includegraphics[width=\hsize]{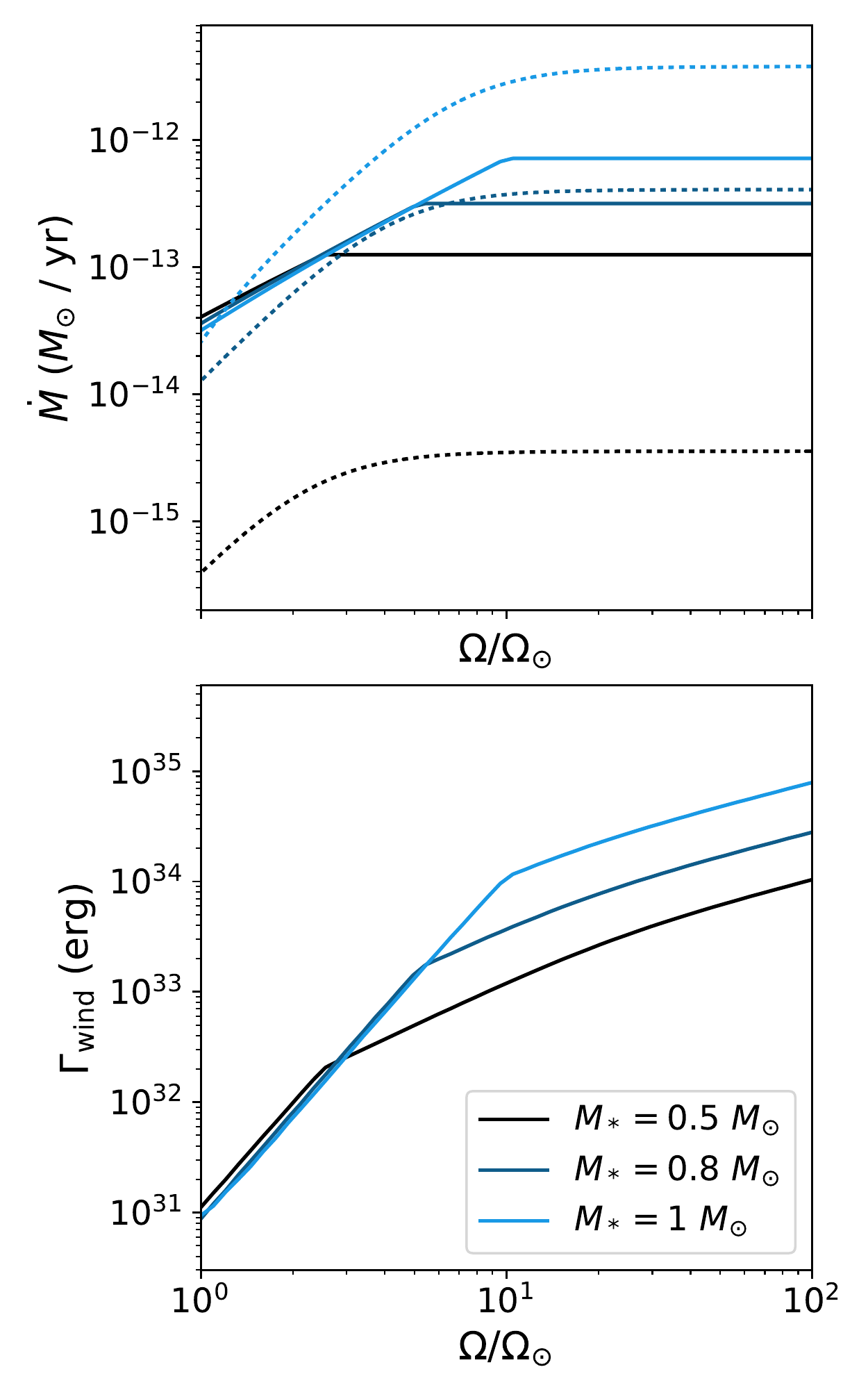}
\caption{Mass loss rate and braking torque as functions of rotational velocity for a star of mass $0.5\ M_{\odot}$ (black), $0.8\ M_{\odot}$ (dark blue) and $1\ M_{\odot}$ (light blue). Top panel: comparison between the mass loss rate obtained with our model (solid curves) and the prescription of \citet{CranmerSaar2011} (dotted curves). Bottom panel: braking torque.}
\label{fig:mass_loss_and_torque}
\end{figure}

Fig.~\ref{fig:mass_loss_and_torque} shows mass loss rate and braking torque as functions of rotational velocity for several stellar masses. As can be seen in the upper panel, the mass loss rate increases with rotation until saturation. This is similar to the behavior of density and temperature at the base of the corona. At high rotation, a dependence on stellar mass appears because stars have different saturation velocities. Despite the fact that our model predicts a mass-loss rate similar to that of \citet{CranmerSaar2011} for a solar-mass star at solar rotation rate, our results suggest a weaker dependence on stellar mass and rotation rate. This discrepancy stems in the additional constraint we put on the spin-down timescale. \citet{Barnes2010} and \citet{Mattetal2015} suggested that this quantity is decreasing with stellar mass for saturated rotators and increasing for unsaturated ones (see section~\ref{subsec:spin_down_time}). Since we aimed at reproducing this behavior, we had to impose a different mass-loss rate from that of \citet{CranmerSaar2011}.

The lower panel shows the braking torque resulting from our calculations. As for mass loss rate, the unsaturated and saturated regimes are clearly visible. The dependence on rotation can be approximated by two power laws:
\begin{align}
\Gamma_{\mathrm{wind}} &\propto \Omega_c & (\mbox{saturated}) \label{eq:sat_torque}\\
\Gamma_{\mathrm{wind}} &\propto \Omega_c^{p+1} & (\mbox{unsaturated}), \label{eq:unsat_torque}
\end{align}
where $p = 2.11$. These exponents are close to those found by \citet{Kawaler1988}. We point that the dependence on mass is negligible in the unsaturated regime and becomes important in the saturated regime. At ten times the solar rotation rate, the torque is one order of magnitude higher for a $1 \Msun$ star than for its $0.5 \Msun$ counterparts.

\subsection{Spin-down time}
\label{subsec:spin_down_time}

From these properties of the stellar coronae, we can now discuss the characteristic timescale of magnetic braking. \citet{Barnes2010} showed that its dependence on mass was not the same for fast and slow rotators. Among the former, more massive stars tend to spin down more quickly whereas this trend reverses for slow rotators. \citet{Mattetal2015} defined two different spin-down times for the saturated and unsaturated regimes and reproduced this result. To investigate the mass dependence predicted by ESPEM, we calculated the equivalent expression of these timescales as functions of the angular momentum of the star and the braking torque:
\begin{align}
\tau_{\mathrm{sat}} &= \frac{I_* \Omega_c}{\Gamma_{\mathrm{wind}}} & (\mbox{saturated}) \\
\tau_{\mathrm{unsat}} &= \frac{I_* \Omega_c}{p \Gamma_{\mathrm{wind}}} \left(\frac{\Omega_c}{\Omega_{\odot}}\right)^p & (\mbox{unsaturated}),
\end{align}
where $I_*$ is the stellar moment of inertia. Relations \eqref{eq:sat_torque} and \eqref{eq:unsat_torque} indicate that, with these definitions, $\tau_{\mathrm{sat}}$ and $\tau_{\mathrm{unsat}}$ do not depend on rotational velocity. Indeed, in both saturated and unsaturated regimes, the braking torque increasing with rotation rate compensates the explicit dependence on $\Omega_c$. Thus, they allow a comparison of spin-down timescale for different stellar masses, different ages and the same state of evolution.

\begin{figure}
\centering
\includegraphics[width=\hsize]{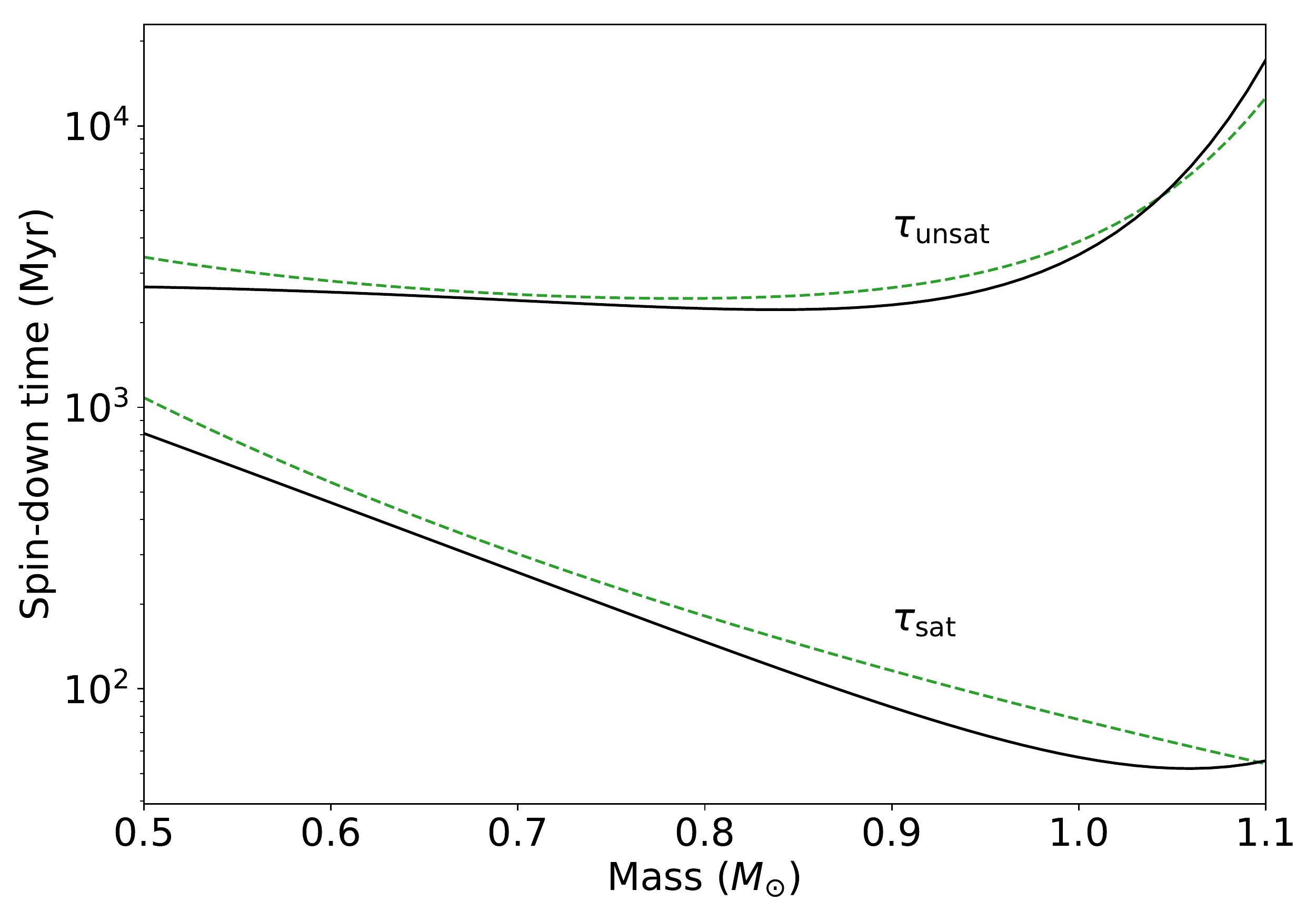}
\caption{Spin-down timescale as a function of stellar mass in the saturated (lower lines) and unsaturated regimes (upper lines). Comparison between values from \citet{Mattetal2015} (dashed green lines) and the starAML routine from \citet{Revilleetal2015b} (solid black lines).}
\label{fig:spin_down_time}
\end{figure}

Fig.~\ref{fig:spin_down_time} shows the spin-down timescale as a function of stellar mass for both saturated and unsaturated regimes. The slopes of both curves are determined by the competition between two quantities increasing with stellar mass: the moment of inertia on the one hand, and the braking torque on the other hand. Since the latter does not vary significantly with mass for slow rotators, the corresponding timescale increases proportionaly to the moment of inertia. On the contrary, for fast rotators, this competition results in a spin-down timescale decreasing with mass ($\tau_{\mathrm{sat}} \propto M_*^{-3.86}$). The figure also shows a very good agreement of our results with the model of \citet{Mattetal2015}.

\subsection{Secular evolution}
\label{subsec:SeculEvol}

We applied this braking torque to stellar evolutionary tracks for masses $0.5 M_{\odot}$, $0.8 M_{\odot}$ and $1 M_{\odot}$ computed with the STAREVOL code \citep{Amardetal2016} to assess the variations of the rotational velocity of solar-type stars on secular timescales. For each star, we considered three different initial conditions, a slow, median and fast rotator, corresponding respectively to the 25th percentile, median, and 90th percentile of the observed distribution of rotational velocities of young stars. We used the parameters given by \citet{GalletBouvier2015} to set the free parameters of our model: initial rotation period, coupling constant and disk lifetime (see Table~\ref{tab:internal_constant}).

\begin{figure}
\centering
\includegraphics[width=\hsize]{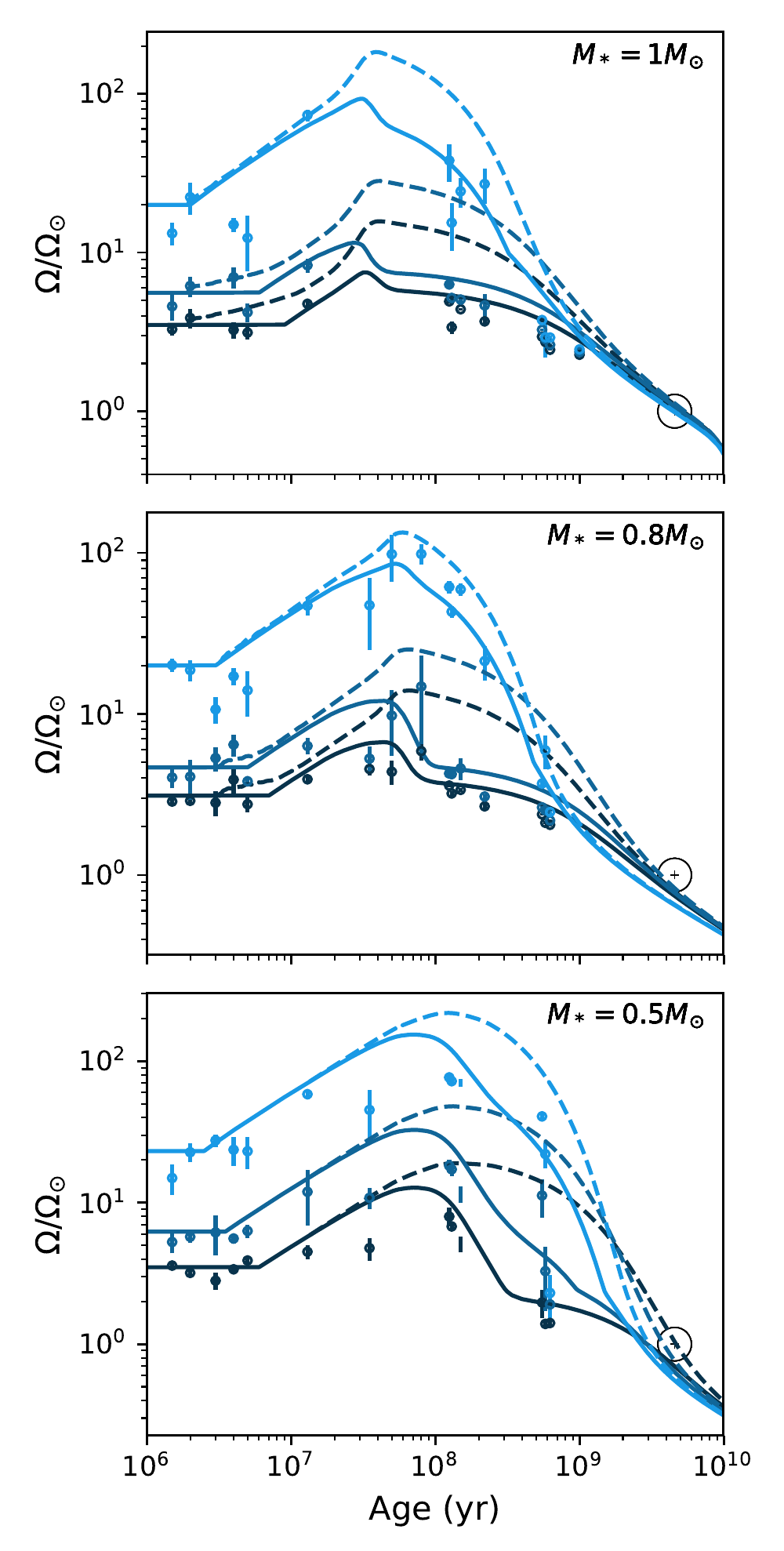}
\caption{Secular evolution of the rotation for stars of mass $1 M_{\odot}$, $0.8 M_{\odot}$ and $0.5 M_{\odot}$ and different initial conditions. Solid and dashed curves represent the rotation velocity of the envelope and the core, respectively. The solar velocity at solar age is pictured by the white circle in the bottom right corner of each frame. The blue circles with error bars correspond to the 25th, 50th and 90th percentiles of rotational distributions of observed stellar clusters. These values were published by \citet{GalletBouvier2015}. Values for the internal coupling constant $\tint$ are given in Table~\ref{tab:internal_constant}.}
\label{fig:secular_evolution}
\end{figure}

\begin{table}
\caption{Values of the internal coupling constant $\tint$ used to compute the secular evolution in Fig.~\ref{fig:secular_evolution}. These values were found by \citet{GalletBouvier2015} who fitted a semi-analytical model on observations of stellar rotation.}
\label{tab:internal_constant}
\centering
\begin{tabular}{l c c c}
\hline\hline
Stellar mass  & $0.5\ \Msun$ & $0.8\ \Msun$ & $1\ \Msun$ \\
\hline
Slow rotator   & 500 Myr & 80 Myr & 30 Myr \\
Median rotator & 300 Myr & 80 Myr & 28 Myr \\
Fast rotator   & 150 Myr & 15 Myr & 10 Myr \\
\hline
\end{tabular}
\end{table}

The rotation of the envelope was kept constant from the beginning of the simulation until a time corresponding to the dissipation of the disk. During this phase, the radiative zone is free to spin up under contraction. However, it is linked to the external layer of the star through internal coupling. After the disk dissipation, rotational velocities of both envelope and core evolve under the action of stellar evolution, coupling and magnetic braking, as described in section~\ref{sec:model}.

Fig.~\ref{fig:secular_evolution} shows the evolution of rotation of stars of masses $0.5 M_{\odot}$, $0.8 M_{\odot}$ and $1 M_{\odot}$ from the PMS to the end of the MS. For each mass, the initial spread in rotational velocities is reduced over the MS and all stars converge on a unique sequence. Our model predicts that rotation of evolved stars is proportional to $t^{-0.473}$ in accordance with the Skumanich law \citep{Skumanich1972}. As is visible in the top panel, at the age of the Sun, solar-mass stars rotate at the solar rate. Less massive stars are slower at this age, which had been found by \citet{GalletBouvier2015}. This result is also in agreement with those of \citet{Mattetal2015}, who noted that the braking timescale in the unsaturated regime is shorter for lower mass stars.

\section{Tidal dissipation}
\label{sec:tides}

In this section, we detail how the tidal torque $\Gtide$ is computed. We work within the tidal quality factor formalism \citep{Goldreich1963,Kaula1964,MacDonald1964}. In this framework, the response of the star to tidal perturbation is measured by the modified tidal quality factor, $Q'$, which is the ratio of the total energy stored in the tidal velocity field divided by the energy dissipated over one revolution of the planet. This convenient formalism has been used to account for stellar structure and rotation variations over the secular evolution of the system \citep{Mathis2015,BolmontMathis2016,GalletCharbonnelAmardetal2017}. \citet{MathisLePoncinLafitte2009} showed that, when the planet is far enough ($a > 5 \ R_*$) and weakly deformed, it could be considered as a point-mass disturber for calculating tidal dissipation within the star. In this case, the deformation of the star due to gravitational perturbation is quadrupolar and only the corresponding frequency is excited: $\omega_{\mathrm{tide}} = 2 \ (\Omega_c - n)$ in the circular coplanar case. As the planet gets closer to its star, which corresponds to a semi-major axis inferior to 0.018 AU in the case of a solar-radius star, this hypothesis is no longer valid. However, since the planets orbiting their star below this limit are destroyed after a few thousand years in our calculations, we do not take this complication into account in this work. Nevertheless, differences arising from the fact that the planet is an extended body should be studied in a future work. Tidal frequency depends on the rotation rate of the envelope because we only consider the dissipation in the convective envelope of the star. The torque depends on the dissipation rate and the structural parameters of the star and the planet \citep{MurrayDermott1999}:
\begin{equation}
\Gamma_{\mathrm{tide}} = - \mathrm{sign} (\omega_{\mathrm{tide}}) \frac{9}{4 Q'} \frac{G m_{\mathrm{p}}^2}{a^6} R_*^5,
\label{eq:tidal_torque}
\end{equation}
where $m_{\mathrm{p}}$ is the mass of the planet, assumed to be constant over the evolution of the system. We interpolated the STAREVOL evolutionary tracks \citep{Amardetal2016} to obtain the values of stellar structure at each ESPEM time step. Since tidal dissipation originates from two different physical mechanisms, the equilibrium tide and the dynamical tide, each of them have to be studied separately. The total tidal dissipation is proportional to the inverse of the equivalent quality factor. It is the sum of the two contributions:
\begin{equation}
\frac{1}{Q'} = \frac{1}{Q_{\mathrm{eq}}'} + \frac{1}{Q_{\mathrm{dyn}}'},
\label{eq:tidal_quality_factor}
\end{equation}
where $Q_{\mathrm{eq}}'$ and $Q_{\mathrm{dyn}}'$ are the modified quality factor related to the equilibrium and the dynamical tide, respectively.

\subsection{Equilibrium tide}

To compute the equivalent quality factor related to the equilibrium tide, we used the observational results of \citet{Hansen2012}, who calibrated the value of the dissipation for stellar masses ranging from 0.3 to 1.5 $M_{\odot}$. He worked in the constant time lag framework \citep{Mignard1979,Hut1981,Eggletonetal1998}, which is different from ours. Therefore, we had to reconcile both formalisms. The equilibrium tide corresponds to the hydrostatic adjustment of the star to the perturbation of the planet. In this work, we identify it to its quadrupolar moment. In the adiabatic case, the axis of the tidal bulge would be aligned with the one joining the centers of the star and of the planet. However, dissipation induces an angle $2 \delta$ between these axes. In the quality factor framework, dissipation is proportional to this angle and $Q'$ is related to the lag angle through the relation $2 \delta = 3 / (2 Q')$. The time lag $\Delta \tau = 2 \delta / |\omtide|$ corresponds to the time delay between the positions of the bulge and the axis joining the centers. In the constant time lag framework, dissipation is quantified by the constant $\sigma_*$, which measures the ratio of energy loss due to tidal friction divided by the magnitude of the quadrupolar moment of the deformation. \citet{Hansen2012} found an empirical law between the normalized dissipation $\bar{\sigma}_* = \sigma_* / \sigma_0$, where $\sigma_0 = \sqrt{ G / (M_{\odot} R_{\odot}^7)}$, and the stellar parameters:
\begin{equation}
\bar{\sigma}_* = \Delta \tau \frac{G}{\sigma_0 R_*^5}.
\end{equation}

Following \citet{BolmontMathis2016}, we inverted the relation to obtain the corresponding modified tidal quality factor:
\begin{equation}
\Qeq = \frac{3}{2} \frac{1}{\sigma_0\ \bar{\sigma}_*} \frac{G}{R_*^5\ |\omega_{\mathrm{tide}}|}.
\label{eq:equivalence_Q_sigma}
\end{equation}

We point that the quality factor associated with the equilibrium tide is inversely proportional to tidal frequency. Thus, we expect different predictions than those of models assuming a constant quality factor \citep{ZhangPenev2014}. This dependence might cause computational problems close to co-rotation. In practice, we replaced $|\omega_{\mathrm{tide}}|$ in equation~\eqref{eq:equivalence_Q_sigma} by $|\omega_{\mathrm{tide}}| + \varepsilon_{\mathrm{tide}}$ where $\varepsilon_{\mathrm{tide}} = 10^{-10}\ \mathrm{rad.s}^{-1}$. This technique was introduced by \citet{BolmontMathis2016}, who showed that this does not affect the predictions for the systems' final states. Moreover, since $\Gtide \propto \mathrm{sign}(\omtide) / \Qeq$, i. e. $\Gtide \propto \mathrm{sign}(\omtide) \omtide$ with Eq.~\eqref{eq:tidal_torque}, there is no actual singularity of the torque here.

\subsection{Dynamical tide}

At low frequency, the time-varying gravitational potential excites inertial waves in the convective envelope of the rotating star. This phenomenon occurs for $\omega_{\mathrm{tide}} \in [-2\ \Omega_c,\ 2\ \Omega_c]$ and constitutes the wavelike part of the tidal response. Its dissipation may be several orders of magnitude larger than that of the equilibrium tide. \citet{OgilvieLin2007} first calculated the dissipation of the dynamical tide within solar-type stars' interiors. Their result is highly dependent on tidal frequency because of resonances induced by the wavelike nature of the dynamical tide. Moreover, the properties of the resonances strongly depend on the assumed values of the eddy coefficient applied to tidal waves to account for the turbulent friction applied by convection. These latter are still poorly known, as pointed by e. g. \citet{OgilvieLesur2012,OgilvieLin2007,Gueneletal2016} and \citet{Mathisetal2016}. This would imply heavy calculations to study secular orbital evolution with complex hydrodynamical simulation for each step of the structural and rotational evolution of the star \citep{WitteSavonije2002,Auclair-Desrotouretal2014}. To address this situation, \citet{Ogilvie2013} computed the frequency-averaged value of the tidal waves dissipation using an impulsional method. Over the range $[-2\ \Omega_c,\ 2\ \Omega_c]$, this leads to a result that depends only on the structure and rotation of the star. \citet{Mathis2015} then applied this method to the envelopes of solar-like stars to compute the equivalent modified quality factor associated with the dynamical tide:
\begin{multline}
\dfrac{3}{2 Q_{\mathrm{dyn}}'} = \dfrac{100 \pi}{63} \epsilon^2 \left(\dfrac{\alpha^5}{1 - \alpha^5}\right) (1 - \gamma)^2 (1 - \alpha)^4 \times \\
\dfrac{(1 + 2\alpha + 3\alpha^2 + \tfrac{3}{2} \alpha^3)\left[1 + \left(\tfrac{1 - \gamma}{\gamma}\right) \alpha^3 \right]}{\left[1 + \tfrac{3}{2}\gamma + \tfrac{5}{2\gamma} (1 + \tfrac{1}{2}\gamma - \tfrac{3}{2}\gamma^2) \alpha^3 - \tfrac{9}{4}(1 - \gamma)\alpha^5\right]^2}. \label{eq:dissipation}
\end{multline}
In this formula, $\alpha = R_{r} / R_*$ and $\beta = M_r / M_*$ are the mass and aspect ratios of the core, respectively, $\gamma = \alpha^3 (1 - \beta) / (\beta (1 - \alpha^3))$ is the ratio of the envelope's density over that of the core and $\epsilon = \Omega / \sqrt{G M_{\odot} / R_{\odot}^3}$. We computed these ratios from the stellar evolution tracks of the STAREVOL code \citep{Amardetal2016}, which is well adapted to our bi-layer stellar rotation model. \citet{Mathis2015} showed that, for a given rotation rate, the dissipation could vary over several orders of magnitude with aspect and mass ratios. Moreover, a solar-like star rotation rate may vary during its life on the MS in the range $[\Omega_{\odot},\ 100 \ \Omega_{\odot}]$ \citep{GalletBouvier2015}. Since the dissipation is proportional to the rotation rate squared, a given star may dissipate significantly more at its arrival on the MS than at solar age, as discussed by~\citet{GalletBolmontMathisetal2017}. In this paper, we focused on stars with solar metallicity. The impact of this parameter should be carefully studied in a further work, as done by \citet{Bolmontetal2017}, who showed that tidal dissipation in the convective zone of solar-like stars was higher for metal poor stars on the PMS and that this trend was inversed on the MS.

\begin{figure}
\centering
\includegraphics[width=\hsize]{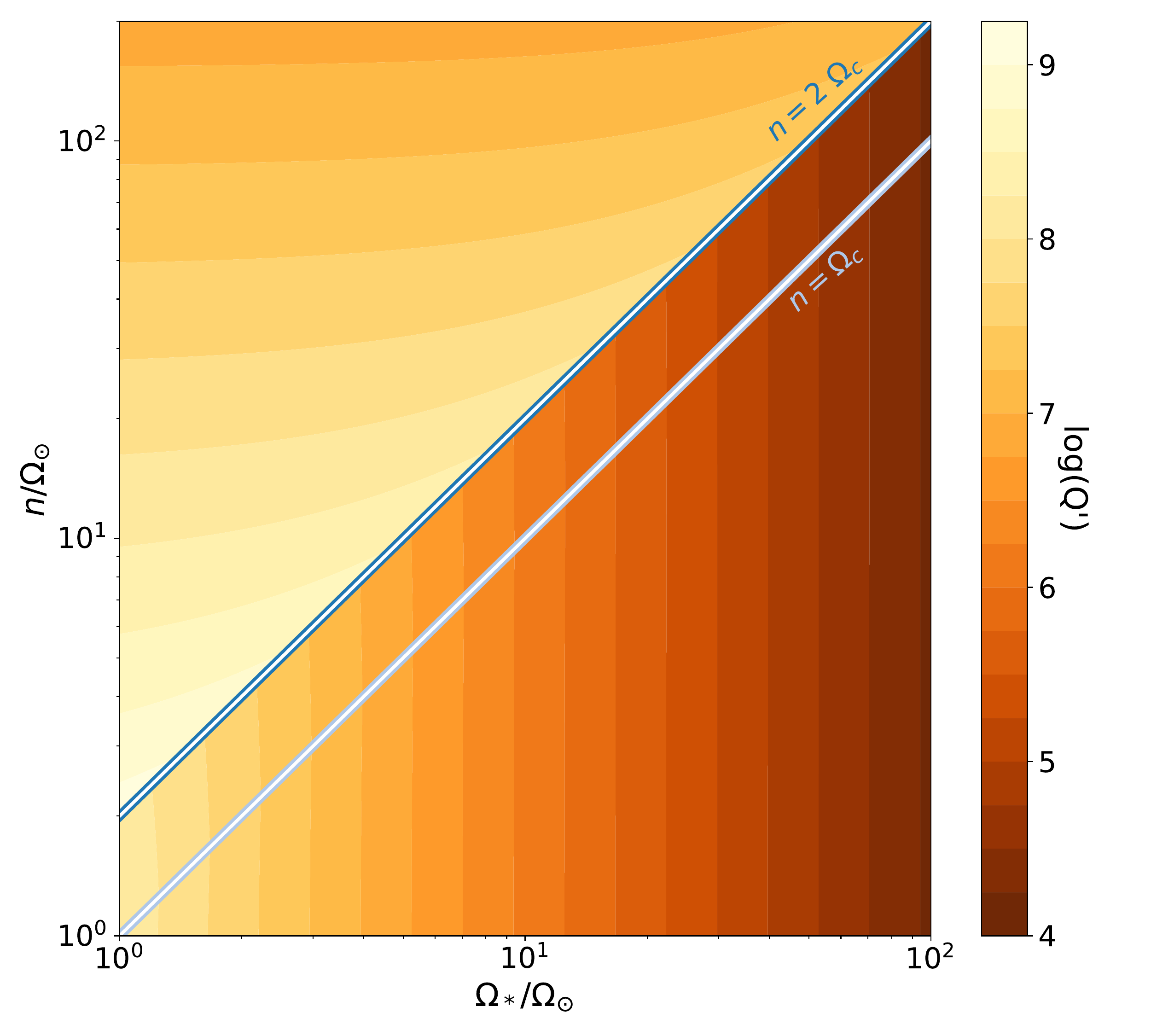}
\caption{Equivalent tidal quality factor as a function of stellar rotation and orbital mean motion for a solar-mass star. The white line with dark blue contour represents the limit ($n = 2 \Omega_c$) below which tidal inertial waves propagate in the envelope. The white line with light blue contour ($n = \Omega_c$) corresponds to co-rotation. Above this line, planets are migrating inward. Below, they are pushed away from the star.}
\label{fig:quality_Om_n}
\end{figure}

Fig.~\ref{fig:quality_Om_n} shows the tidal quality factor $Q'$ from Eq.~\eqref{eq:tidal_quality_factor} for a star of mass $1 M_{\odot}$ at solar age as a function of rotation rate and orbital mean motion. It is visible that tidal dissipation is the sum of two contributions. The white line with dark blue contour delimits the domain of application of inertial waves. Below, the tidal frequency is in the range $[-2\ \Omega_c,\ 2\ \Omega_c]$ because $n < 2\ \Omega_c$. In this domain, the dissipation of the dynamical tide dominates that of the equilibrium tide. It is visible on the figure that the dynamical tide contributes to significantly enhance the tidal dissipation. Indeed, the quality factor in the region where they are raised is orders of magnitude lower than in the upper part of the figure. The dependence of the dynamical tide on rotation also clearly appears. Indeed, in the lower region, the contour lines are vertical since the dependence on tidal frequency of the dissipation of the dynamical tide was averaged in this work. On the contrary, the contour lines in the upper region are actually determined by $\omtide$ (see Eq.~\ref{eq:equivalence_Q_sigma}).

This figure also reveals three possible regimes of tidal interaction. Below the light blue line, the orbital motion is slower than the stellar rotation rate. This region corresponds to planets beyond the co-rotation radius. They are pushed away as a consequence of the dissipation of the equilibrium and dynamical tides. Between the blue lines, the dynamical tide applies but planets are below the co-rotation radius. Consequently, they spiral inward under the effect of both the equilibrium and dynamical tides. Above the dark blue line, inertial waves are no longer raised. In this region, planets spiral inward under the sole influence of the dissipation of the equilibrium tide. The secular evolution of a star-planet system generally involves the succession of different phases during which the system is in one of these three states.

\section{Results}
\label{sec:results}

In this section, we analyze the coupled influence of magnetic braking and tidal dissipation on the secular evolution of star-planet systems. We begin by showing that the rotation of a star can be significantly impacted by the presence of a planet, especially when the latter falls into the former. The fate of a star-planet system can follow very different scenarios, depending on parameters such as the star's mass and initial rotation rate, the planetary mass, and the initial semi-major axis of the orbit. To investigate these dependences, we browse the parameter space to study which combinations lead to the planet's demise. We then define a criterion allowing to say, for a given system, if the planet will survive or be destroyed.

\subsection{Impact of a planet on its host star's rotation rate}

We first investigated the consequences of tidal dissipation in the convective envelope on stellar rotation. To that end, we computed the secular evolution of a star-planet system composed by a solar-mass star and a Jupiter-mass planet with an initial semi-major axis equal to 0.025 AU. In order to compare the results with those of Fig.~\ref{fig:secular_evolution}, we calculated this evolution for three different initial rotation periods: 1.4, 5 and 8 days.

\begin{figure}
\centering
\includegraphics[width=\hsize]{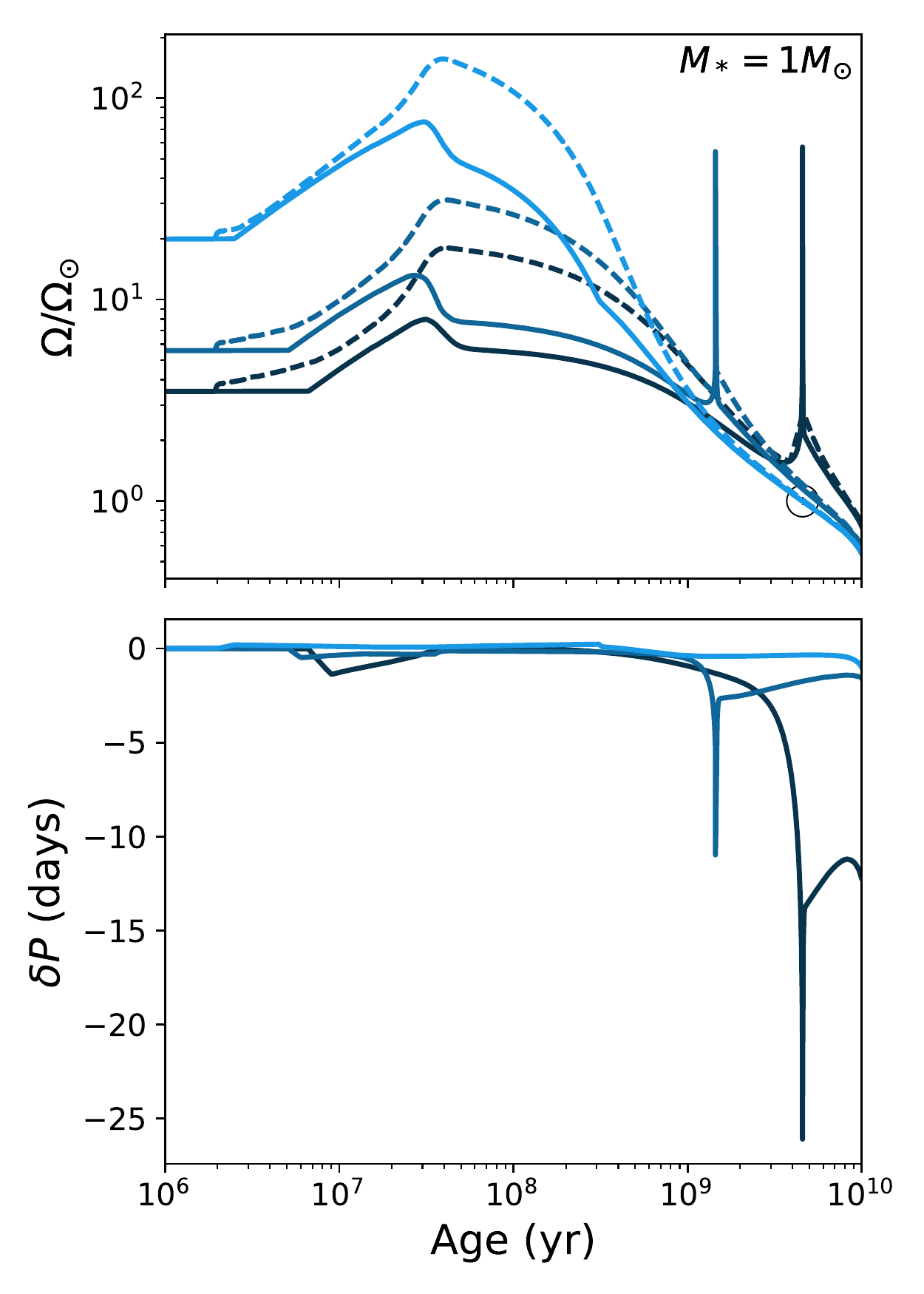}
\caption{\textit{Top panel:} Secular evolution of a star of mass 1 $\Msun$ orbited by a planet of mass 1 $\Mjup$ with initial semi-major axis $\aini$ = 0.025 AU. From light to dark blue: $\Pini$ = 1.4, 5, 8 days. For each color, the solid line represents the rotation of the envelope and the dashed one, that of the core. \textit{Bottom panel:} Difference in rotation period to the reference case of Fig.~\ref{fig:secular_evolution} in absolute value. The color scheme is the same as for the top panel.}
\label{fig:first_res_fig}
\end{figure}

As visible in Fig.~\ref{fig:first_res_fig}, the stellar rotation can be strongly impacted by the presence of a planet. The most noticeable differences with Fig.~\ref{fig:secular_evolution} are the two peaks in the slow rotator's (dark blue) and median rotator's (medium blue) curves. They are both caused by the destruction of the planet. Indeed, as will be further analysed later in this section, when the planet orbits the star below its co-rotation radius, the orbit gives angular momentum to the stellar spin, causing the central body to spin up and the planet to migrate closer to the star. This process is maintained until either the semi-major axis crosses the co-rotation radius, in which case the sense of the tidal exchange of angular momentum is reversed, or the planet is disrupted after having approached too close to the star. In the latter case, the angular momentum transfer from the orbit to the stellar spin during the inward spiral motion of the planet is important enough to increase the rotation rate of the envelope by one order of magnitude. After the planet has been destroyed, the wind brakes the envelope rapidly so that a peak is formed in the rotation rate evolution curve.

The bottom panel shows the difference $\delta P$ between the stellar rotation period in the case where the star hosts a planet and the case where it does not:
\begin{equation}
\delta P = P_{*,\ \mbox{with planet}} - P_{*,\ \mbox{without planet}}. \label{eq:defDeltaP}
\end{equation}
This difference quantifies the influence of the planet on stellar rotation. For the medium- and dark-blue curves, it is negative because the star has been spun up. We note that the rotation remains fast after the planet destruction. For the dark-blue curve, $\delta P$ is of the order of 12 days for the last five gigayears of the simulation, which means that the star spins twice faster than what gyrochronology predicts. In the case of the light-blue curve, the star has been spun down because its planet was initially located beyond its corotation radius. This is why $\delta P$ is positive. Since the effect in this case is significantly smaller than in the other two simulations, the scale of the plot does not allow for a detailed analysis of this curve. We refer the reader to Appendix~\ref{subsec:compare_BM} for detailed analysis of positive-$\delta P$ curves.

The fast rotator (light blue curve) seems unaffected by the presence of the planet. Indeed, contrary to the two other simulations, the planet in this case starts orbiting the star beyond its co-rotation radius. This leads to a fast outward migration which has weak effects on the stellar spin. This phenomenon, further discussed in~\ref{subsubsec:secularBrowsing_Pini}, illustrates that the evolution of a star-planet system strongly depends on the parameters characterizing it. In the following, we study how the secular evolutions of the stellar rotation and the orbit differ when varying the initial conditions and parameters.

\subsection{Planet survival time for a solar-mass star}
\label{subsec:tPlanetSurvival}

We now investigate the influence of the system's characteristics on its evolution. We identified four main parameters for this study: the mass of the star, its initial rotation, the mass of the planet and the initial semi-major axis of the orbit. In this subsection, we only consider 1 $\Msun$ stars and focus on how the fate of the system differs when the three other parameters vary. We first analyse the influence of each characteristic individually by comparing  simulations done with different values of this parameter. Then, we study how their crossed variations impact the fate of the system by computing, for a large set of parameters, the time at which the planet is destroyed.

\subsubsection{Initial semi-major axis}
\label{subsubsec:secularBrowsing_aini}

We started by looking at the impact of the initial semi-major axis on the fate of a system. To that end, we computed the secular evolution of a star-planet couple constituted by a 1 $\Msun$ star with an initial rotation period $P_{*, \mathrm{ini}}$ = 5 days and a 1 $\Mjup$ mass planet for three different values of the semi-major axis: 0.025, 0.035, and 0.045 AU. In the following, we compare the results obtained in each case and discuss the differences observed from one scenario to another.

\begin{figure}
\centering
\includegraphics[width=\hsize]{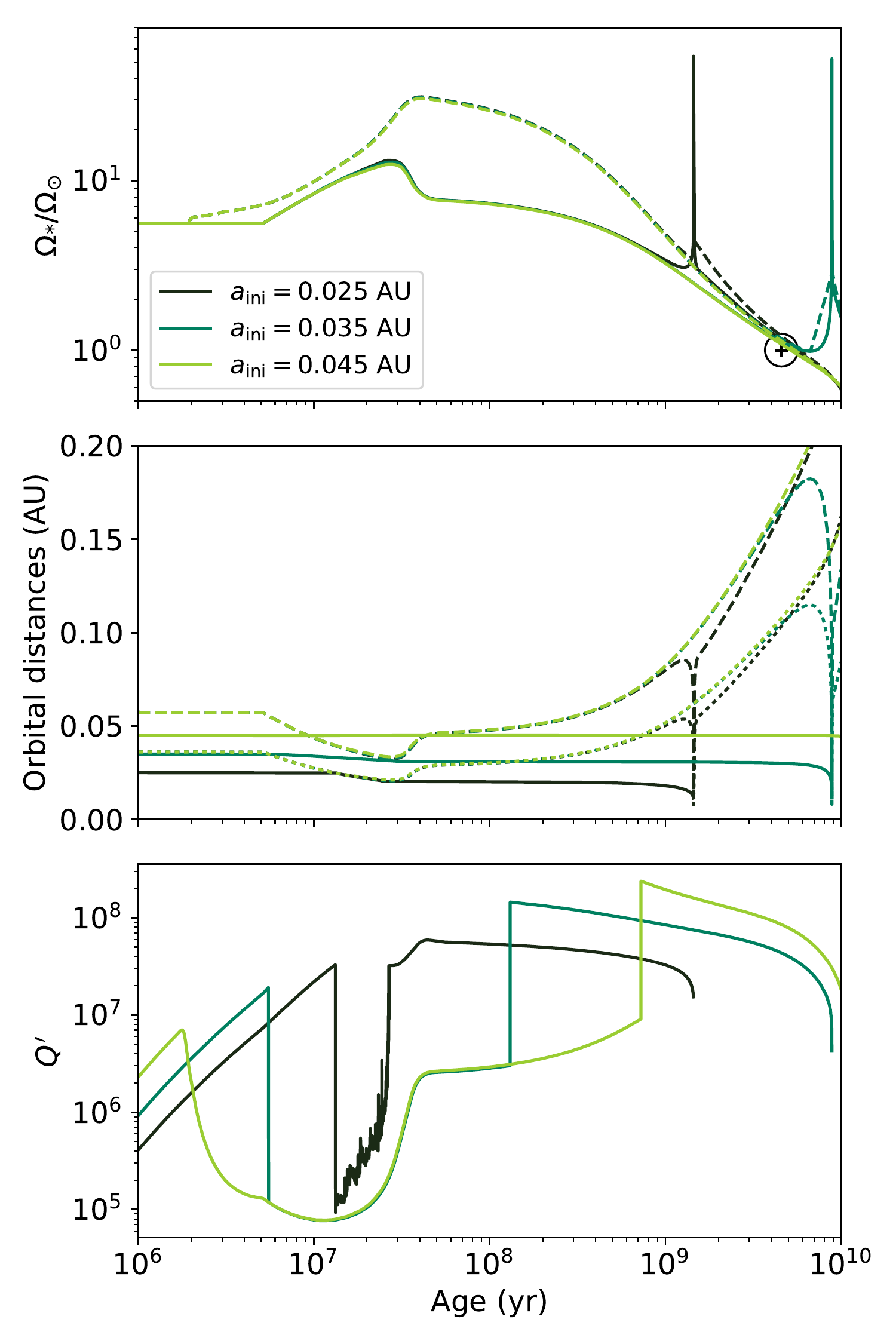}
\caption{Secular evolution of a star-planet system with $M_*$ = 1 $\Msun$, $\Pini$ = 5 days and $m_{\mathrm{p}}$ = 1 $\Mjup$. The initial semi-major axis was set to 0.025 AU (dark green curves), 0.035 AU (medium green curves) and 0.045 AU (light green curves). \textit{Top panel:} Rotation rate of the envelope (solid lines) and of the radiative zone (dashed lines) of the star. \textit{Middle panel:} semi-major axis (solid lines), co-rotation radius of the star (dashed lines) and limit of excitation of the dynamical tide (dotted lines). \textit{Bottom panel:} modified tidal quality factor.}
\label{fig:secularBrowsing_aini}
\end{figure}

It appears on Fig.~\ref{fig:secularBrowsing_aini} that the farthest planet (light green curves) is too far from the star and not heavy enough to undergo a significant migration. The closest planet (dark green curves) starts orbiting its star at a distance where only the equilibrium tide is raised. The orbital evolution in this case arises from the variations of the limit of excitation of inertial waves, which is represented by the dotted dark green line in the middle panel of the figure (confounded with the other two dotted curves during the first billion year). As the star spins up during its contraction, this limit decreases until it equals the semi-major axis. Then, the system achieves a resonance state in which $n = 2\ \Omega_c$. This resonance, characterized on the figure by the solid dark green curve sticking to the dotted dark green curve, is well visible in the bottom panel. Indeed, during this phase, the curve representing the quality factor is noisy, which is the consequence of the dynamical tide being alternatively excited and switched off in the convective envelope. This state is maintained until the end of the contraction and has for consequence to significantly move the planet closer to the central body within a short timescale (a few million years) compared to usual tidal evolution timescales. On the main sequence, the planet undergoes orbital evolution due to dissipation of the equilibrium tide and eventually spirals inward to its destruction. The planet in the middle (medium green curves) experiences a similar scenario. However, unlike the previous system, it does not undergo a resonance at the limit of application of inertial waves. This is because the dynamical tide is not as strong as in the former case. As in the case with $\aini$ = 0.025 AU, the planet with $\aini$ = 0.035 AU moves progressively closer to its star during the main sequence until its demise. Figure~\ref{fig:secularBrowsing_aini} shows that the initial semi-major axis directly influences the speed of tidal evolution: when it decreases, the system evolves faster.

\subsubsection{Planetary mass}
\label{subsubsec:secularBrowsing_mp}

We now show how the mass of the planet influences the system. To achieve this, we performed a similar calculation to that of the previous paragraph. We computed the evolution of a star-planet couple with a solar-mass star initially rotating with a period $\Pini$ = 5 days, orbited by a planet at an initial distance $\aini$ = 0.035 AU for three different planetary masses: 0.1, 1 and 5 $\Mjup$. The system with the Jupiter-mass planet is actually the same as the one in the previous paragraph for which $\aini$ = 0.035 AU. Thus, paragraphs~\ref{subsubsec:secularBrowsing_aini} and \ref{subsubsec:secularBrowsing_mp} correspond to variations around the same point in different directions of the parameter space.

\begin{figure}
\centering
\includegraphics[width=\hsize]{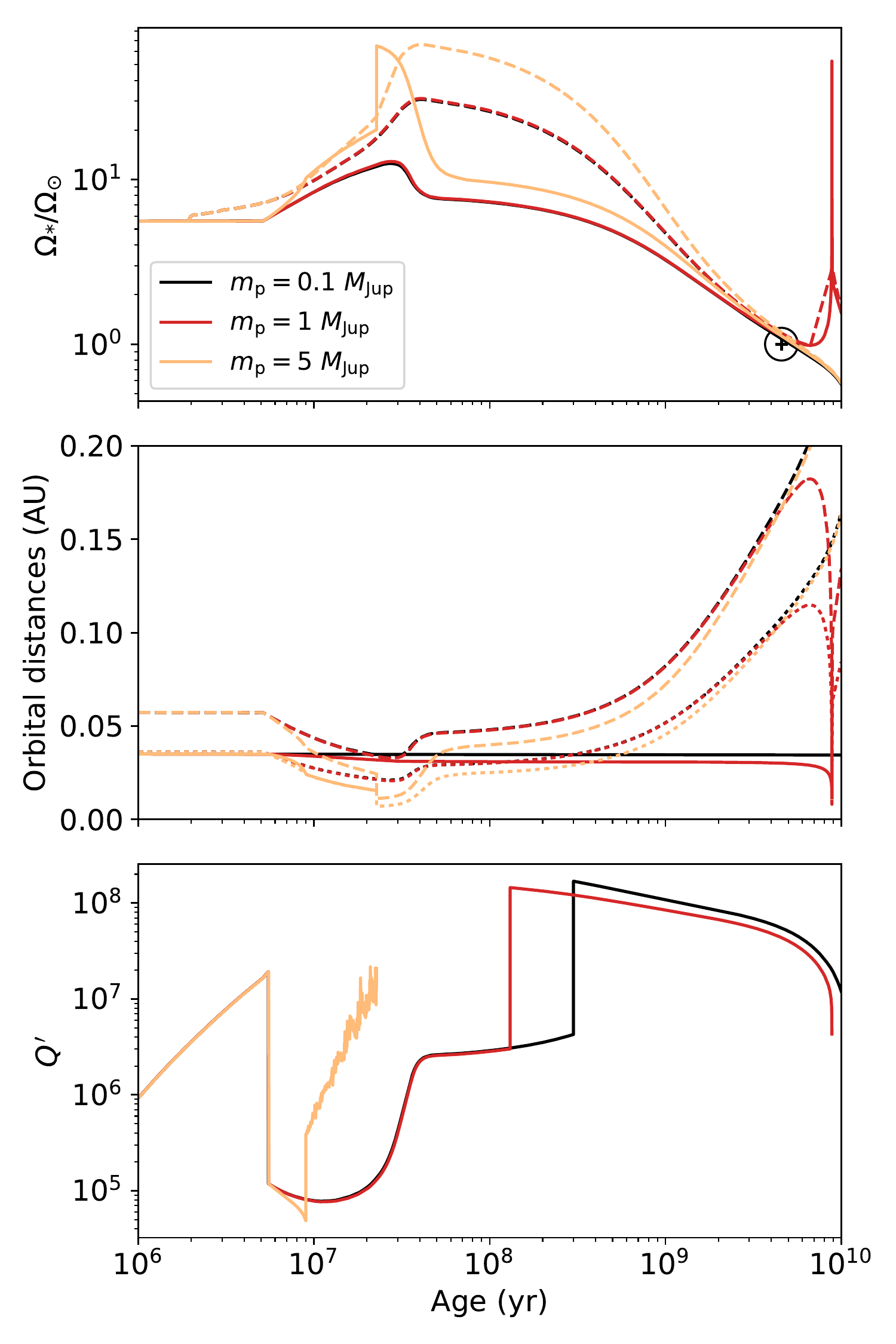}
\caption{Secular evolution of a star-planet system formed by a 1 $\Msun$ star with $\Pini$ = 5 days, orbited by a planet at initial distance $\aini$ = 0.035 AU, for three different values of the planet's mass: $m_{\mathrm{p}}$ = 0.1 $\Mjup$ (black lines), 1 $\Mjup$ (red) and 5 $\Mjup$ (orange). \textit{Top panel:} Rotation rate of the envelope (solid lines) and of the radiative zone (dashed lines) of the star. \textit{Middle panel:} semi-major axis (solid lines), co-rotation radius of the star (dashed lines) and limit of excitation of the dynamical tide (dotted lines). \textit{Bottom panel:} modified quality factor.}
\label{fig:secularBrowsing_mp}
\end{figure}

Figure~\ref{fig:secularBrowsing_mp} shows the evolution of three star-planet systems for a fixed semi-major axis and the varying planetary mass. Similarly to the farthest planet in Fig.~\ref{fig:secularBrowsing_aini}, the lightest planet (black lines) is not enough influenced by tidal dissipation to undergo significant orbital evolution. The heaviest planet (orange lines) is disrupted during the early phases of the system's life. As the semi-major axis of its orbit crosses the limit of excitation of inertial waves, the tidal torque is strong enough to raise a resonance between the orbit and the stellar spin. Unlike the closest planet of Fig.~\ref{fig:secularBrowsing_aini}, the planet here is massive enough to significantly spin up its host star, which results in a decrease of the dynamical tide limit, represented by the orange dotted line in the middle panel. The resonance then imposes the planet to follow this evolution and migrate closer to the star. Thus, the semi-major axis of the orbit collapses in a few million years and the planet is destroyed. The medium-mass planet here corresponds to the same system as the middle planet of Fig.~\ref{fig:secularBrowsing_aini} (the one with $\aini$ = 0.025 AU). Its evolution is an intermediate case of the two previous cases. Figure~\ref{fig:secularBrowsing_mp} shows that tidal evolution is faster for systems with more massive planets. Systems with massive planets behave like systems with close planets since they evolve faster than others. However, a massive planet may spin up its star, which is not possible for a lighter and closer planet.

\subsubsection{Initial stellar rotation rate}
\label{subsubsec:secularBrowsing_Pini}

\begin{figure}
\centering
\includegraphics[width=\hsize]{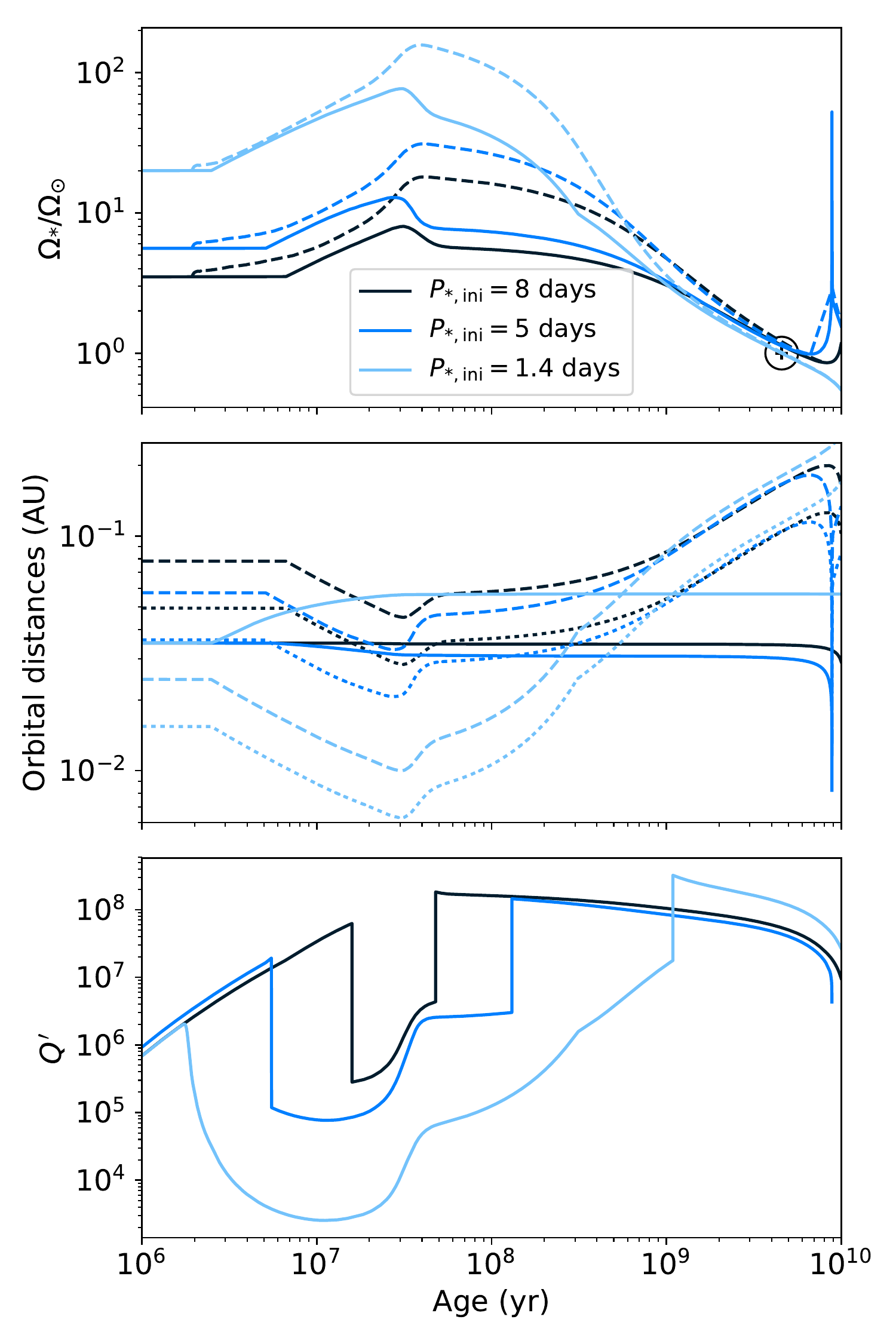}
\caption{Secular evolution of a star-planet system with $M_*$ = 1 $\Msun$, $\aini$ = 0.035 AU and $m_{\mathrm{p}}$ = 1 $\Mjup$. The initial stellar rotation period was set to 8 $\mathrm{days}$ (dark blue lines), 5 $\mathrm{days}$ (blue lines) and 1.4 $\mathrm{days}$ (light blue lines). \textit{Top panel:} Rotation rate of the envelope (solid lines) and of the radiative zone (dashed lines) of the star. \textit{Middle panel:} semi-major axis (solid lines), co-rotation radius of the star (dashed lines) and limit of excitation of the dynamical tide (dotted lines). \textit{Bottom panel:} modified quality factor.}
\label{fig:secularBrowsing_Pini}
\end{figure}

We finally reproduced the calculations of paragraphs~\ref{subsubsec:secularBrowsing_aini} and~\ref{subsubsec:secularBrowsing_mp} to study the influence of the initial stellar rotation rate. We set the planetary mass to 1 $\Mjup$ and the initial semi-major axis to $a_{\mathrm{ini}} = 0.035$ AU and computed the evolution of a system formed by this planet and a solar-mass star with an initial rotation period $P_{*,\mathrm{ini}}$ = 1.4, 5 and 8 days, which correspond to the fast, median and slow rotators defined by \citet{GalletBouvier2015}. The results of these simulations are shown on Fig.~\ref{fig:secularBrowsing_Pini}.

In the system with the fastest rotating star, the planet starts its orbit beyond the co-rotation radius and is rapidly pushed away from the star as a consequence of the dissipation of the dynamical tide. When the star spins down during the main sequence phase, the co-rotation radius increases and eventually exceeds the semi-major axis of the orbit. Consequently, the tidal torque changes signs but, at this distance, tidal dissipation is not strong enough to cause significant orbital evolution and the planet eventually survives. The system with the median rotator, shown in regular blue, corresponds to the same system as the middle planet from Fig.~\ref{fig:secularBrowsing_aini} and the medium-mass planet from Fig.~\ref{fig:secularBrowsing_mp}. Its evolution has been discussed in the previous paragraphs. It is expected that higher initial stellar rotation rates lead to longer planet lifetimes, if we consider the planet lifetime as the duration before it spirals towards its host star. However, the planet orbiting the slow rotator, shown in dark blue, lives longer than the one orbiting the median rotator. This can be explained by considering the contraction phase of the star. For the slow rotator, the limit of excitation of inertial waves is farther away from the star than for the median rotator. Consequently, inertial waves, which result in an enhanced tidal dissipation, are raised over a longer duration in the latter case. Moreover, inertial waves in the median rotator are excited at a higher rotation rate than in the slow rotator. This leads to lower quality factors in the former case. To synthesize, the dynamical tide is raised over a longer duration and induces higher tidal torques in the median rotator, which is why its planet is destroyed before that orbiting the slow rotator. The results of Fig.~\ref{fig:secularBrowsing_Pini} emphasize the impact of the contraction phase on the rest of the orbital evolution. Unlike the two precedent parameters, $m_{\mathrm{p}}$ and $a_{\mathrm{ini}}$, the initial stellar rotation rate does not have a monotonic influence on the planet lifetime.

\subsubsection{Browsing the parameter space}

Determining how the system's characteristics influence the planet lifetime is crucial to achieve a better understanding of orbital evolution. This is why we repeated the previous calculations on a wider and more detailed sample. We set the stellar mass to $1 \ M_{\odot}$ and considered three initial rotation periods: 1.4, 5 and 8 days. For each period, we computed the secular evolution of systems with varying planetary mass and initial semi-major axis. For the former, the bin was equally spaced in logarithm with ten points between 10 $M_{\oplus}$ and 10 $M_{\mathrm{Jup}}$. For the latter, the bin was linearly spaced between 0.02 AU and 0.05 AU. We used the parameters of Table 2 from \citet{GalletBouvier2015} for the disk lifetime and internal coupling constant. For each simulation, we computed the planet lifetime as either the duration before the star terminates its main sequence phase in the case the planet survives, or the duration before the planet is disrupted in the other case.

\begin{figure*}
\centering
\includegraphics[width=17cm]{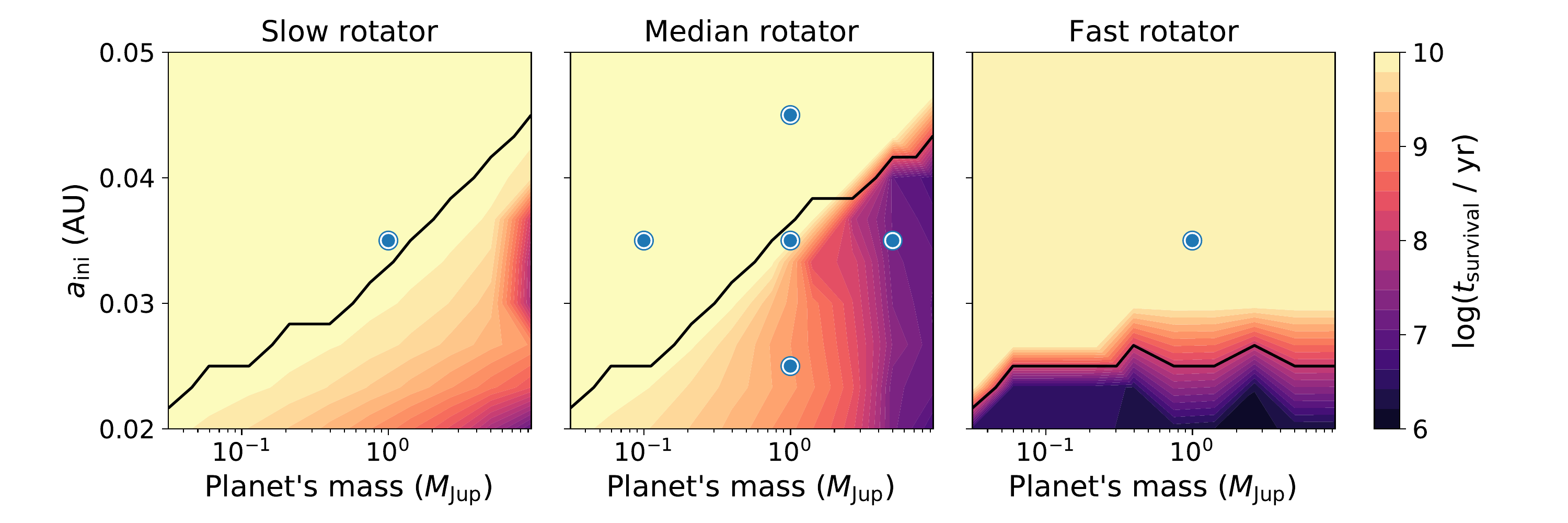}
\caption{Planet lifetime as a function of planetary mass (X axis) and initial semi-major axis (Y axis) for three different initial stellar rotation rates. The color represent the logarithm of the planet lifetime in years (from 6 to 10). On each panel, the thick black line indicates the limit above which planets survive to their host star. \textit{Left panel:} slow rotator, $P_{*,\mathrm{ini}} = 8$ days. The blue dot corresponds to the slow rotator of Fig.~\ref{fig:secularBrowsing_Pini}. \textit{Middle panel:} median rotator, $P_{*,\mathrm{ini}} = 5$ days. The blue dots correspond to the cases treated above in the secular evolution figures~\ref{fig:secularBrowsing_aini},~\ref{fig:secularBrowsing_mp} and~\ref{fig:secularBrowsing_Pini}. \textit{Right panel:} fast rotator, $P_{*,\mathrm{ini}} = 1.4$ days. The blue dot corresponds to the fast rotator of Fig.~\ref{fig:secularBrowsing_Pini}.}
\label{fig:survivalTime1Msun}
\end{figure*}

Fig.~\ref{fig:survivalTime1Msun} shows the planet lifetime as a function of planetary mass and initial semi-major axis for three different initial stellar rotation rates. Its value spans from a few million years (planet decayed during the PMS) to 10 billion years (terminal age on the main sequence). As can be seen in the figure, in this two-bodies simplified approach, for this mass range, planets which formed above 0.045 AU are very likely to survive to their host star. These results confirm the monotonic dependence of the planet lifetime on planetary mass and initial semi-major axis: apart from an outlier for massive close-in planets in the left panel, colors always become darker in the bottom right corner, i.e., closer and heavier planets are destroyed earlier. The blue dots represent the situation on these maps of the cases treated in detail in figures~\ref{fig:secularBrowsing_aini},~\ref{fig:secularBrowsing_mp} and~\ref{fig:secularBrowsing_Pini}.

The thick black line represents the demarcation between the region where planets survive and the one where they eventually spiral inward to their demise. It behaves differently for the fast rotator than for the two other cases. Indeed, in the right panel, the line is horizontal, which means that the survival of the planet does not depend on its mass, whereas in the others, it is oblique, which indicates that the survival of the planet results from a trade-off between its mass and distance from the host star.

This different behavior can be explained by considering the co-rotation radius associated with the initial stellar rotation period, as was analyzed in paragraph~\ref{subsubsec:secularBrowsing_Pini}. For the slow and the median rotators, its value is higher than 0.05 AU, meaning that all planets shown on the left and middle panels of Fig.~\ref{fig:survivalTime1Msun} start their orbit below the co-rotation radius of their host star. On the contrary, the initial co-rotation radius of the fast rotator is about 0.025 AU. As shown in the right panel of Fig.~\ref{fig:survivalTime1Msun}, all planets that start orbiting the fast rotator below this limit are rapidly destroyed, leading to a disentangling of the population in two groups: planets which formed above the co-rotation radius, rapidly pushed away from their stars, which allowed them to eventually survive, and those which formed below and were disrupted within the first million years.

This implies a depopulation of the region close to fast rotating stars, as observed by \citet{McQuillanMazehAigrain2013}. \citet{TeitlerKonigl2014}, who used a secular evolution code to explain this dearth, concluded that it was due to tidal engulfment. Our results suggest it has actually two causes: tidal disruption for the closest planets and fast outward migration in the early stages of the evolution for the other, as shown for instance in Fig.~\ref{fig:secularBrowsing_Pini}. This second alternative that could solve the problem was pointed by \citet{LanzaShkolnik2014}. They found that neither the tidal quality factor framework, nor the constant viscous time model \citep{Eggletonetal1998}, could reproduce the distribution of planetary orbital periods and favored a scenario in which close-in planets form at a distance of about 1 AU from their star and migrate closer to their star because of chaotic dynamic evolution caused by the other planets of the system.

In the cases of the slow and median rotators, the thick line involves more complex, intricate dependences on the characteristics of the system. The slope of the line in the $(m_{\mathrm{p}},\ a_{\mathrm{ini}})$ plane results from the competition between the planetary mass and semi-major axis in the expression of the tidal torque. This competition is analyzed in the following subsection.

\subsection{Survival rate criterion}

The aim of this subsection is to determine a criterion allowing to predict whether the planet will survive knowing the characteristics of the system ($M_*$, $P_{*,\mathrm{ini}}$, $m_{\mathrm{p}}$ and $a_\mathrm{ini}$), which we will refer to as the survival rate criterion. To that end, we repeated the simulations of Fig.~\ref{fig:survivalTime1Msun} for stars of mass 0.5, 0.6, ..., 1.1 $M_{\odot}$. For each star, we defined a slow, median and fast rotator as \citet{GalletBouvier2015}. Then, we calculated the planet lifetime as a function of $m_{\mathrm{p}}$ and $a_\mathrm{ini}$ for each rotator. Finally, we defined two regions, the survival region, where planets survive to their host star, and the disruption region, where planets are eventually disrupted.

\begin{figure*}
\centering
\includegraphics[width=17cm]{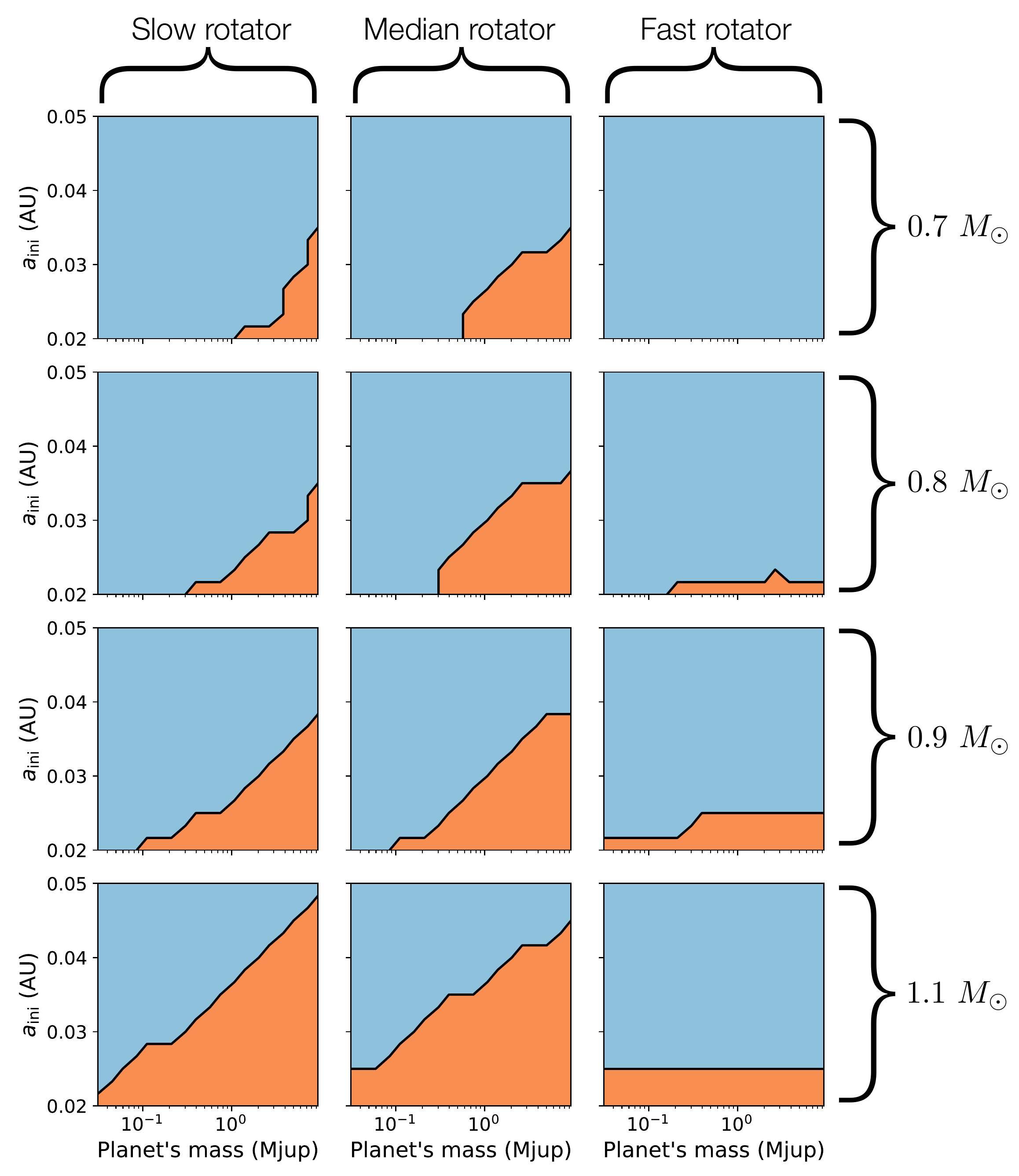}
\caption{Survival and disruption regions for stars of mass 0.7, 0.8, 0.9 and 1.1 $M_{\odot}$. Each row corresponds to a given stellar mass, each column to a given initial stellar rotation. Slow, median, and fast rotators correspond respectively to the 25th, 50th, and 90th percentiles of stellar rotation distributions observed in open clusters as defined by \citet{GalletBouvier2015}. For each panel, the orange area represents the region of the ($m_{\mathrm{p}},\ a_{\mathrm{ini}}$) plane where planets are eventually destroyed and the blue area, the region where planets survive.}
\label{fig:planetSurvivalAllMasses}
\end{figure*}

Fig.~\ref{fig:planetSurvivalAllMasses} shows these regions for various stellar masses and initial stellar rotations. On each panel, the orange region indicates the planetary masses and initial semi-major axes for which the planet does not survive to tidal interaction. It is always located in the bottom right corner, which illustrates that, the more a planet is massive and close to its host star, the more it is likely to be eventually decayed. The figure also emphasizes the radically different behavior of fast rotators. Indeed, in the right column, the line delimiting the blue and orange regions is horizontal, suggesting as in Fig.~\ref{fig:survivalTime1Msun} that the survival of the planet does not depend on its mass. For the median and slow rotators, behaviors are similar and the border between the two regions is a straight line whose slope is the same for all stellar masses and initial rotation rates. The decayed-planets region is wider for higher mass stars, which is in agreement with the result of \citet{BolmontMathis2016} who found that tidal evolution has a stronger impact on systems with higher mass stars despite the fact that the bulk dissipation is less important.

In each panel of Fig.~\ref{fig:planetSurvivalAllMasses}, the thick black line separates the image into two half spaces. Getting the equation of this border as a function of the characteristics ($M_*$, $P_{*,\mathrm{ini}}$, $m_{\mathrm{p}}$ and $a_{\mathrm{ini}}$) is sufficient to define the survival rate criterion. For initially fast rotating stars, the survival of the planet is determined by its initial semi-major axis. If the latter is greater than the co-rotation radius, the planet survives; in the other case it is rapidly decayed. For other initial stellar rotations, the other parameters play a more significant role in the system's fate. It is visible in Fig.~\ref{fig:secularBrowsing_Pini} that the initial stellar rotation does have a clear influence on the survival of the planet. Therefore, we sought a relation involving only $M_*$, $m_{\mathrm{p}}$ and $a_{\mathrm{ini}}$ to determine the equation of the border. For each stellar mass, we collected the coordinates of the points on the border in the slow and median rotator cases. Since increasing (resp. decreasing) the stellar or planetary mass (resp. the semi-major axis) reduces the chances of planet survival, we expected that all these points arrange according to the following equation:
\begin{equation}
\frac{M_* m_{\mathrm{p}}^{\beta}}{a_{\mathrm{ini}}^{\alpha}} = K, \label{eq:survival_surface}
\end{equation}
where $\alpha$, $\beta$ and $K$ are constants that we fitted with our simulated data.

\begin{figure}
\centering
\includegraphics[width=\hsize]{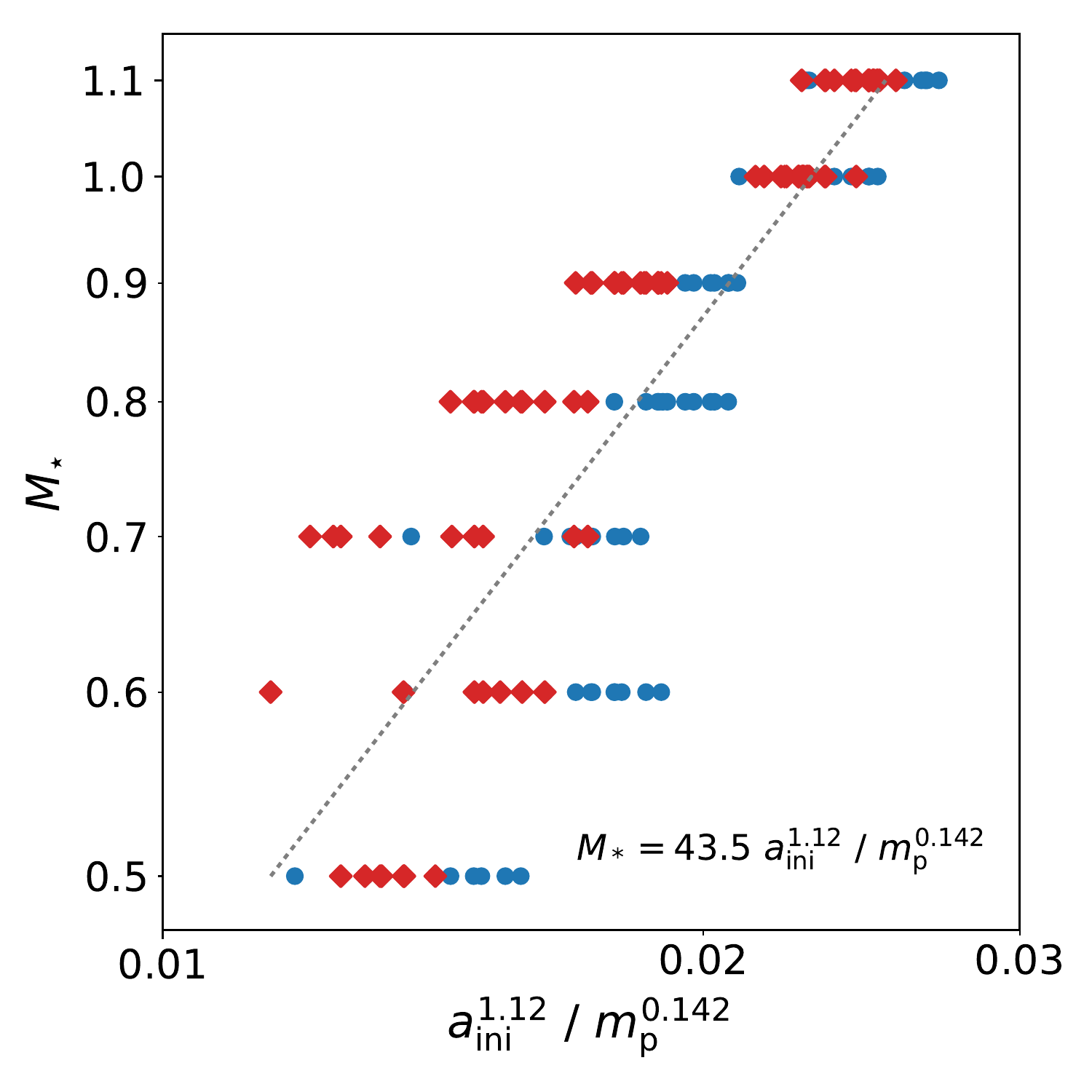}
\caption{Coordinates of the points of the border between the survival and demise regions (respectively the blue and orange regions of Fig.~\ref{fig:planetSurvivalAllMasses}) for varying stellar mass (in units of $\Msun$), planetary mass (in units of $\Mjup$) and initial semi-major axis (in AU). The blue circles correspond to the median rotator case and the red diamonds to the slow rotator. The dotted gray line represents the result of our fit (Eq.~\ref{eq:survival_surface}).}
\label{fig:staying_alive_criterion}
\end{figure}

Fig.~\ref{fig:staying_alive_criterion} shows the result of our fit of the border between the survival and disruption regions. The stellar mass is expressed in solar masses, the planetary mass in jovian masses and the semi-major axis in astronomical units. The points are scattered because the relation between the parameters on the border is not an ideal power law. Despite this dispersion, a trend clearly appears in the figure. We used a multidimensional least squares method based on the Levenberg-Marquardt algorithm to fit the results of our simulations and found that the points follow the law $M_* = K\ \aini^{\alpha}\ /\ m_{\mathrm{p}}^{\beta}$ with $K = 43.5$, $\alpha = 1.12$ and $\beta = 0.142$. This equation allows us to define the survival rate criterion:
\begin{align}
\mathrm{If}\ \frac{M_* m_{\mathrm{p}}^{\beta}}{a_{\mathrm{ini}}^{\alpha}} < K,& \mbox{ then the planet survives.}\\
\mathrm{If}\ \frac{M_* m_{\mathrm{p}}^{\beta}}{a_{\mathrm{ini}}^{\alpha}} > K,& \mbox{ then it is destroyed.}
\end{align}

\section{Conclusion}
\label{sec:conclusion}

We presented ESPEM, a code implementing a model of secular evolution of star-planet systems under magnetic braking and tidal interaction. Our wind model based on the work of \citet{Revilleetal2015a} allows a fine analysis of magnetic phenomena occurring in the stellar corona. Its results are in agreement with recent observations \citep{Barnes2010,GalletBouvier2015} and has the advantage of explaining the mass dependence of the braking torque \citep[see][]{Mattetal2015}. We emphasized the difference between the ab-initio tidal prescription of \citet{Mathis2015} and the constant quality factor model used by \citet{ZhangPenev2014}. The former may predict that a planet will survive whereas the latter predicts its decay. Thus, understanding the physical processes at stake in tidal interaction is of prime importance to correctly interpret the various architectures of planetary systems discovered during the last decade. We showed some examples of complex orbital evolution scenarios that could occur as a consequence of tidal dissipation within the star. We showed that the spin-down caused by a planet orbiting a fast rotator could be detected only for planets more massive than Jupiter. We showed that, for hot Jupiters orbiting solar-like stars at a distance superior to 0.05 AU ($\sim 10.7\ R_{\odot}$), tidal interaction had no significant impact on the system. Finally, we defined a survival rate criterion, allowing to predict the fate of a star-planet system (survival or decay of the planet) knowing its main parameters.

It is now of prime importance to consider the impact of  tides within the stellar radiative zone, which are constituted by gravity waves and dissipated by turbulent friction or breaking \citep[see e. g.][]{Guillotetal2014}. Indeed, this phenomenon is likely to enhance the tidal torque or compete with the dissipation in the envelope and modify the rotation rate of the stellar radiative core. Thus, it is necessary to take it into account to build a consistent model of tidal evolution. The present work can be considered as a step towards this aim. Implementing a physical description of internal angular momentum transport in the star is also among the perspectives of this paper \citep[e. g.][and references therein]{Zahn1992,Mathis2013}. Even if the simplified two-layer model implemented in ESPEM is already able to give good insights on stellar rotation evolution, a more detailed view of the interior as developed by \citet{Amardetal2016} and \citet{GalletCharbonnelAmardetal2017} would deeply improve our comprehension.

After the physical phenomena in the star are better treated in the code, considering tidal dissipation within the planet will be the next step. Indeed, whereas the spin angular momentum of the planet is by several orders of magnitude smaller than that of the star, tides raised in the former may lead to comparable tidal torques \citep[see][and references therein]{MathisRemus2013,Ogilvie2014}. Such theoretical studies could be compared to the recent results of \citet{OconnorHansen2018}, who studied hot-Jupiter populations to constrain the dissipation in such planets. Taking them into account is required to study planet spin synchronization and alignment and orbital circularization and migration in the early stages of the system's life. The frequency dependence of the dynamical tide \citep{Auclair-Desrotouretal2014}, as well as the impact of eccentricity, inclination and dynamical instabilities arising in multi-planet systems, have to be taken into account to compare theoretical predictions with the observed distributions \citep[e. g.][]{Bolmontetal2015,DamianiMathis2018}.

We will also consider the consequences of Roche-lobe overflow in a future work. Indeed, this process may cause mass-transfer from the planet to the star which would result in an outward migration to conserve angular momentum \citep{Trillingetal1998}. In their theoretical study, \citet{Guetal2003} showed that this scenario occurs in eccentric systems. More recently, \citet{Jacksonetal2017} developed a consistent model of Roche-lobe overflow and showed that some observed systems were experiencing significant mass transfer. Such developments will be useful to improve our model of star-planet interaction.

In this work, we only considered stars with dipolar magnetic fields. However, the strength and topology of the global surface magnetic field of a star evolves with its age and structure \citep{Gregoryetal2012,Vidottoetal2014,Emeriau-ViardBrun2017}. This may have an impact on the wind and stellar spin-down during the main sequence \citep{BrunBrowning2017,Metcalfeetal2016}. Convection is able to influence magnetism via dynamo processes \citep{Brunetal2004, Brunetal2015, Brownetal2011, Strugareketal2017Sci, Augustsonetal2015, Kapylaetal2014} and differential rotation \citep{Brunetal2017}. Determining the impact of these mechanisms is necessary to ensure realistic rotational evolutions of stars.

This work also opens a discussion on the temperature and density at the base of the corona of solar-like stars (see Eqs.~\ref{eq:corDensity} and \ref{eq:corTemperature}). An empirical relation between the coronal temperature, the mass, and the stellar rotation rate was proposed by \citet{JohnstoneGudel2015}, who studied the correlation between the coronal temperature and the X-ray flux of low-mass main-sequence stars. From scaling laws and the relation between the X-ray luminosity and the stellar rotation period found by \citet{Reinersetal2014}, they derived the expression $T_c \propto M^{-0.42}\ \Omega_*^{0.52}$ which differs from what we used \citep{HolzwarthJardine2007}. Concerning the density at the base of the corona, the hypothesis of \citet{HolzwarthJardine2007} was justified by coronal electron density measurements from \citet{IvanovaTaam2003}. Both the reliablity of these observations and the possibility of inferring the base density of the wind from such quantities were criticized \citep[on the former point, see][]{Gudel2004}. However, more recent stellar wind models do not contradict this law \citep{CranmerSaar2011,Suzukietal2013}. In a future work, we will consider changing the assumptions we made on coronal parameters for relations taking into account the stellar structure and evolution. However, such developments are beyond the scope of the present paper.

Another important point of improvement is the Rossby number. In this paper, we used the expression developed by \citet{CranmerSaar2011} for their models of main-sequence solar-like stars to express the Rossby number and used the value of $\Teff$ at the ZAMS in the formulation. However, convective turnover times are significantly longer for young stars than for their main-sequence counterparts. Thus, even slowly-rotating pre-main-sequence stars may be in saturated regime, which has been observed by recent X-ray surveys on young clusters \citep{Getmanetal2005,Gudeletal2007}. This suggests that only stars older than a certain age can reach the unsaturated regime. The limit age has to be of the order of 13 Myr, the age of the h Per cluster which was studied by \citet{Argiroffietal2016}. They found that it was the youngest known cluster that behaves as clusters of main-sequence stars.

Recent works started to account for the variations of the Rossby number with stellar structure using stellar evolution tracks \citep{Folsometal2016,SadeghiArdestanietal2017}. Both computed the convective turnover time at each time step as the ratio of the local pressure scale height $H_p$ divided by the local convective velocity at one $H_p$ over the base of the convective zone. These approaches are close to that of \citet{Spadaetal2013}, who published evolutionary tracks of solar-like stars for various compositions and mixing length parameters, allowing to study the variations of the convective turnover timescale along stellar evolution. The peculiarity of these formulations is that they allow consistent treatment of the Rossby number over the PMS phase, which to our knowledge is not taken into account in other secular evolution models of rotation of solar-like stars. The question of the Rossby number was also extensively studied by \citet{Brunetal2017}. In particular, they highlighted that the fluid Rossby number $\Rof$, defined as the ratio of the advection term divided by the Coriolis force, was a good indicator to assess the large-scale rotational behavior of solar-like stars. Based on numerical simulations performed with the ASH code, they gave an expression of this number as a function of stellar mass and surface rotation rate.

In a future work, we will take into account the studies mentioned above to improve this part of our model. A better understanding of the magnetic and coronal properties of solar-like stars will also be useful in order to implement star-planet magnetic interactions in the code \citep{Strugareketal2014,Strugareketal2015}. Recently, \citet{Strugareketal2017} proved that this effect could compete with tidal dissipation in shaping star-planet systems.

Finally, in this work we only considered isolated star-planet systems. Future developments of our code will include implementing dynamical interactions in multiplanet systems \citep[e. g.][]{Laskaretal2012,Bolmontetal2015}. Developing such physical models is important to extract the maximum information from the data of upcoming space missions such as PLATO \citep{Raueretal2014}.

\begin{acknowledgements}
The authors acknowledge Florian Gallet, Emeline Bolmont, Antoine Strugarek, and Rafael A. García for helpful discussion during the project. We also thank Jérémy Ahuir for providing Figs.~\ref{fig:th_test_tidal_limit} and~\ref{fig:th_test_evol}. S.Mathis acknowledges funding by the European Research Council through ERC grant SPIRE 647383. We also thank the ERC STARS2 207430, FP7 SPACEINN, the ANR Toupies SIMI5-6 020 01, the INSU/PNST and CNES support via our PLATO and Solar Orbiter fundings. We thank the referee and editor for useful comments that improved the content of the paper.
\end{acknowledgements}

\bibliographystyle{aa}
\bibliography{biblio}

\begin{thebibliography}{133}
\expandafter\ifx\csname natexlab\endcsname\relax\def\natexlab#1{#1}\fi

\bibitem[{Altschuler \& Newkirk(1969)}]{AltschulerNewkirk1969}
Altschuler, M.~D. \& Newkirk, G. 1969, SoPh, 9, 131

\bibitem[{Amard {et~al.}(2016)Amard, Palacios, Charbonnel, Gallet, \&
  Bouvier}]{Amardetal2016}
Amard, L., Palacios, A., Charbonnel, C., Gallet, F., \& Bouvier, J. 2016, A\&A,
  587, A105

\bibitem[{Anglada-Escudé {et~al.}(2016)Anglada-Escudé, Amado, Barnes,
  Berdiñas, Butler, Coleman, de~La~Cueva, Dreizler, Endl, Giesers, Jeffers,
  Jenkins, Jones, Kiraga, Kürster, López-González, Marvin, Morales, Morin,
  Nelson, Ortiz, Ofir, Paardekooper, Reiners, Rodríguez, Rodríguez-López,
  Sarmiento, Strachan, Tsapras, Tuomi, \& Zechmeister}]{AngladaEscudeetal2016}
Anglada-Escudé, G., Amado, P.~J., Barnes, J., {et~al.} 2016, Nature, 7617, 437

\bibitem[{{Argiroffi} {et~al.}(2016){Argiroffi}, {Caramazza}, {Micela},
  {Sciortino}, {Moraux}, {Bouvier}, \& {Flaccomio}}]{Argiroffietal2016}
{Argiroffi}, C., {Caramazza}, M., {Micela}, G., {et~al.} 2016, \aap, 589, A113

\bibitem[{{Auclair-Desrotour} {et~al.}(2014){Auclair-Desrotour}, {Le
  Poncin-Lafitte}, \& {Mathis}}]{Auclair-Desrotouretal2014}
{Auclair-Desrotour}, P., {Le Poncin-Lafitte}, C., \& {Mathis}, S. 2014, \aap,
  561, L7

\bibitem[{{Augustson} {et~al.}(2015){Augustson}, {Brun}, {Miesch}, \&
  {Toomre}}]{Augustsonetal2015}
{Augustson}, K., {Brun}, A.~S., {Miesch}, M., \& {Toomre}, J. 2015, \apj, 809,
  149

\bibitem[{{Baraffe} {et~al.}(2014){Baraffe}, {Chabrier}, {Fortney}, \&
  {Sotin}}]{Baraffeetal2014}
{Baraffe}, I., {Chabrier}, G., {Fortney}, J., \& {Sotin}, C. 2014, Protostars
  and Planets VI, 763

\bibitem[{{Barker} \& {Ogilvie}(2009)}]{BarkerOgilvie2009}
{Barker}, A.~J. \& {Ogilvie}, G.~I. 2009, \mnras, 395, 2268

\bibitem[{{Barker} \& {Ogilvie}(2010)}]{BarkerOgilvie2010}
{Barker}, A.~J. \& {Ogilvie}, G.~I. 2010, \mnras, 404, 1849

\bibitem[{Barnes(2010)}]{Barnes2010}
Barnes, S. 2010, ApJ, 722, 222

\bibitem[{{Bolmont} {et~al.}(2017){Bolmont}, {Gallet}, {Mathis}, {Charbonnel},
  {Amard}, \& {Alibert}}]{Bolmontetal2017}
{Bolmont}, E., {Gallet}, F., {Mathis}, S., {et~al.} 2017, \aap, 604, A113

\bibitem[{Bolmont \& Mathis(2016)}]{BolmontMathis2016}
Bolmont, E. \& Mathis, S. 2016, CeMDA, 1

\bibitem[{{Bolmont} {et~al.}(2015){Bolmont}, {Raymond}, {Leconte}, {Hersant},
  \& {Correia}}]{Bolmontetal2015}
{Bolmont}, E., {Raymond}, S.~N., {Leconte}, J., {Hersant}, F., \& {Correia},
  A.~C.~M. 2015, \aap, 583, A116

\bibitem[{{Bolmont} {et~al.}(2012){Bolmont}, {Raymond}, {Leconte}, \&
  {Matt}}]{Bolmontetal2012}
{Bolmont}, E., {Raymond}, S.~N., {Leconte}, J., \& {Matt}, S.~P. 2012, \aap,
  544, A124

\bibitem[{{Bouvier}(2013)}]{Bouvier2013}
{Bouvier}, J. 2013, in EAS Publications Series, Vol.~62, EAS Publications
  Series, ed. P.~{Hennebelle} \& C.~{Charbonnel}, 143--168

\bibitem[{Bouvier \& Cébron(2015)}]{BouvierCebron2015}
Bouvier, J. \& Cébron, D. 2015, MNRAS, 453, 3720

\bibitem[{{Brown} {et~al.}(2011){Brown}, {Miesch}, {Browning}, {Brun}, \&
  {Toomre}}]{Brownetal2011}
{Brown}, B.~P., {Miesch}, M.~S., {Browning}, M.~K., {Brun}, A.~S., \& {Toomre},
  J. 2011, \apj, 731, 69

\bibitem[{{Brown}(2014)}]{Brown2014}
{Brown}, T.~M. 2014, \apj, 789, 101

\bibitem[{{Brun} \& {Browning}(2017)}]{BrunBrowning2017}
{Brun}, A.~S. \& {Browning}, M.~K. 2017, Living Reviews in Solar Physics, 14, 4

\bibitem[{Brun {et~al.}(2015)Brun, Browning, Dikpati, Hotta, \&
  Strugarek}]{Brunetal2015}
Brun, A.~S., Browning, M.~K., Dikpati, M., Hotta, H., \& Strugarek, A. 2015,
  SSR, 196, 101

\bibitem[{Brun {et~al.}(2004)Brun, Miesch, \& Toomre}]{Brunetal2004}
Brun, A.-S., Miesch, M.~S., \& Toomre, J. 2004, ApJ, 614, 1073

\bibitem[{{Brun} {et~al.}(2011){Brun}, {Miesch}, \& {Toomre}}]{Brunetal2011}
{Brun}, A.~S., {Miesch}, M.~S., \& {Toomre}, J. 2011, \apj, 742, 79

\bibitem[{{Brun} {et~al.}(2017){Brun}, {Strugarek}, {Varela}, {Matt},
  {Augustson}, {Emeriau}, {DoCao}, {Brown}, \& {Toomre}}]{Brunetal2017}
{Brun}, A.~S., {Strugarek}, A., {Varela}, J., {et~al.} 2017, \apj, 836, 192

\bibitem[{Cranmer \& Saar(2011)}]{CranmerSaar2011}
Cranmer, S.~R. \& Saar, S.~H. 2011, ApJ, 741, 54 (23 pp.)

\bibitem[{Damiani \& Lanza(2015)}]{DamianiLanza2015}
Damiani, C. \& Lanza, A.~F. 2015, A\&A, 574, id. A39, 20 pp.

\bibitem[{{Damiani} \& {Mathis}(2018)}]{DamianiMathis2018}
{Damiani}, C. \& {Mathis}, S. 2018, ArXiv e-prints [\eprint[arXiv]{1803.09661}]

\bibitem[{{Dobbs-Dixon} {et~al.}(2004){Dobbs-Dixon}, {Lin}, \&
  {Mardling}}]{Dobbs-Dixonetal2004}
{Dobbs-Dixon}, I., {Lin}, D.~N.~C., \& {Mardling}, R.~A. 2004, \apj, 610, 464

\bibitem[{Eggleton {et~al.}(1998)Eggleton, Kiseleva, \& Hut}]{Eggletonetal1998}
Eggleton, P.~P., Kiseleva, L.~G., \& Hut, P. 1998, ApJ, 499, 853

\bibitem[{{Emeriau-Viard} \& {Brun}(2017)}]{Emeriau-ViardBrun2017}
{Emeriau-Viard}, C. \& {Brun}, A.~S. 2017, \apj, 846, 8

\bibitem[{{Ferraz-Mello}(2013)}]{Ferraz-Mello2013}
{Ferraz-Mello}, S. 2013, Celestial Mechanics and Dynamical Astronomy, 116, 109

\bibitem[{Ferraz-Mello {et~al.}(2015)Ferraz-Mello, Tadeu~dos Santos, Folonier,
  Czismadia, do~Nascimento, \& Pätzold}]{Ferraz-Melloetal2015}
Ferraz-Mello, S., Tadeu~dos Santos, M., Folonier, H., {et~al.} 2015, ApJ, 807,
  L42

\bibitem[{{Finley} \& {Matt}(2017)}]{FinleyMatt2017}
{Finley}, A.~J. \& {Matt}, S.~P. 2017, \apj, 845, 46

\bibitem[{{Finley} \& {Matt}(2018)}]{FinleyMatt2018ApJ857}
{Finley}, A.~J. \& {Matt}, S.~P. 2018, \apj, 854, 78

\bibitem[{{Folsom} {et~al.}(2016){Folsom}, {Petit}, {Bouvier}, {L{\`e}bre},
  {Amard}, {Palacios}, {Morin}, {Donati}, {Jeffers}, {Marsden}, \&
  {Vidotto}}]{Folsometal2016}
{Folsom}, C.~P., {Petit}, P., {Bouvier}, J., {et~al.} 2016, \mnras, 457, 580

\bibitem[{{Gallet} {et~al.}(2017{\natexlab{a}}){Gallet}, {Bolmont}, {Mathis},
  {Charbonnel}, \& {Amard}}]{GalletBolmontMathisetal2017}
{Gallet}, F., {Bolmont}, E., {Mathis}, S., {Charbonnel}, C., \& {Amard}, L.
  2017{\natexlab{a}}, \aap, 604, A112

\bibitem[{Gallet \& Bouvier(2015)}]{GalletBouvier2015}
Gallet, F. \& Bouvier, J. 2015, A\&A, 577, A98

\bibitem[{{Gallet} {et~al.}(2017{\natexlab{b}}){Gallet}, {Charbonnel}, {Amard},
  {Brun}, {Palacios}, \& {Mathis}}]{GalletCharbonnelAmardetal2017}
{Gallet}, F., {Charbonnel}, C., {Amard}, L., {et~al.} 2017{\natexlab{b}}, \aap,
  597, A14

\bibitem[{Garc\'{i}a {et~al.}(2014)Garc\'{i}a, Ceillier, Salabert, Mathur, van
  Saders, Pinsonneault, Ballot, Beck, Bloemen, Campante, Davies,
  do~Nascimento~Jr., Mathis, Metcalfe, Nielsen, Su\'{a}rez, Chaplin,
  Jim\'{e}nez, \& Karoff}]{Garciaetal2014rotation}
Garc\'{i}a, R.~A., Ceillier, T., Salabert, D., {et~al.} 2014, A\&A, 572, A34

\bibitem[{{Garraffo} {et~al.}(2015){Garraffo}, {Drake}, \&
  {Cohen}}]{Garraffoetal2015}
{Garraffo}, C., {Drake}, J.~J., \& {Cohen}, O. 2015, \apjl, 807, L6

\bibitem[{{Getman} {et~al.}(2005){Getman}, {Flaccomio}, {Broos}, {Grosso},
  {Tsujimoto}, {Townsley}, {Garmire}, {Kastner}, {Li}, {Harnden}, {Wolk},
  {Murray}, {Lada}, {Muench}, {McCaughrean}, {Meeus}, {Damiani}, {Micela},
  {Sciortino}, {Bally}, {Hillenbrand}, {Herbst}, {Preibisch}, \&
  {Feigelson}}]{Getmanetal2005}
{Getman}, K.~V., {Flaccomio}, E., {Broos}, P.~S., {et~al.} 2005, \apjs, 160,
  319

\bibitem[{Gillon {et~al.}(2017)Gillon, Triaud, Demory, Jehi, Agol, Deck,
  Lederer, de~Wit, Burdanov, Ingalls, Bolmont, Leconte, Raymond, Selsis, M.,
  Barkaoui, Burgasser, Burleigh, Carey, Chaushev, Copperwheat, Delrez,
  Fernandes, Holdsworth, Kotze, Van~Grootel, Almleaky, Benkhaldoun, Magain, \&
  Queloz}]{Gillonetal2017}
Gillon, M., Triaud, A. H. M.~J., Demory, B.-O., {et~al.} 2017, Nature, 542, 456

\bibitem[{Goldreich(1963)}]{Goldreich1963}
Goldreich, P. 1963, MNRAS, 126, 257

\bibitem[{{Goldreich} \& {Nicholson}(1989)}]{GoldreichNicholson1989}
{Goldreich}, P. \& {Nicholson}, P.~D. 1989, \apj, 342, 1079

\bibitem[{{Gregory} {et~al.}(2012){Gregory}, {Donati}, {Morin}, {Hussain},
  {Mayne}, {Hillenbrand}, \& {Jardine}}]{Gregoryetal2012}
{Gregory}, S.~G., {Donati}, J.-F., {Morin}, J., {et~al.} 2012, \apj, 755, 97

\bibitem[{{Gu} {et~al.}(2003){Gu}, {Lin}, \& {Bodenheimer}}]{Guetal2003}
{Gu}, P.-G., {Lin}, D.~N.~C., \& {Bodenheimer}, P.~H. 2003, \apj, 588, 509

\bibitem[{{G{\"u}del}(2004)}]{Gudel2004}
{G{\"u}del}, M. 2004, \aapr, 12, 71

\bibitem[{{G{\"u}del} {et~al.}(2007){G{\"u}del}, {Briggs}, {Arzner}, {Audard},
  {Bouvier}, {Feigelson}, {Franciosini}, {Glauser}, {Grosso}, {Micela},
  {Monin}, {Montmerle}, {Padgett}, {Palla}, {Pillitteri}, {Rebull}, {Scelsi},
  {Silva}, {Skinner}, {Stelzer}, \& {Telleschi}}]{Gudeletal2007}
{G{\"u}del}, M., {Briggs}, K.~R., {Arzner}, K., {et~al.} 2007, \aap, 468, 353

\bibitem[{{Guenel} {et~al.}(2016){Guenel}, {Baruteau}, {Mathis}, \&
  {Rieutord}}]{Gueneletal2016}
{Guenel}, M., {Baruteau}, C., {Mathis}, S., \& {Rieutord}, M. 2016, \aap, 589,
  A22

\bibitem[{{Guillot} {et~al.}(1996){Guillot}, {Burrows}, {Hubbard}, {Lunine}, \&
  {Saumon}}]{Guillotetal1996}
{Guillot}, T., {Burrows}, A., {Hubbard}, W.~B., {Lunine}, J.~I., \& {Saumon},
  D. 1996, \apjl, 459, L35

\bibitem[{{Guillot} {et~al.}(2014){Guillot}, {Lin}, {Morel}, {Havel}, \&
  {Parmentier}}]{Guillotetal2014}
{Guillot}, T., {Lin}, D.~N.~C., {Morel}, P., {Havel}, M., \& {Parmentier}, V.
  2014, in EAS Publications Series, Vol.~65, EAS Publications Series, 327--336

\bibitem[{Hairer {et~al.}(2000)Hairer, Nørsett, \& Wanner}]{Haireretal2000}
Hairer, E., Nørsett, S.~P., \& Wanner, G. 2000, Solving Ordinary Differential
  Equations. I - Nonstiff Problems, 2nd ed. (Springer-Verlag Berlin Heidelberg)

\bibitem[{Hansen(2012)}]{Hansen2012}
Hansen, B. M.~S. 2012, ApJ, 757, 6

\bibitem[{{Holzwarth} \& {Jardine}(2007)}]{HolzwarthJardine2007}
{Holzwarth}, V. \& {Jardine}, M. 2007, \aap, 463, 11

\bibitem[{Hut(1981)}]{Hut1981}
Hut, P. 1981, A\&A, 99, 126

\bibitem[{{Ivanova} \& {Taam}(2003)}]{IvanovaTaam2003}
{Ivanova}, N. \& {Taam}, R.~E. 2003, \apj, 599, 516

\bibitem[{{Jackson} {et~al.}(2017){Jackson}, {Arras}, {Penev}, {Peacock}, \&
  {Marchant}}]{Jacksonetal2017}
{Jackson}, B., {Arras}, P., {Penev}, K., {Peacock}, S., \& {Marchant}, P. 2017,
  \apj, 835, 145

\bibitem[{{Johnstone} \& {G{\"u}del}(2015)}]{JohnstoneGudel2015}
{Johnstone}, C.~P. \& {G{\"u}del}, M. 2015, \aap, 578, A129

\bibitem[{{Johnstone} {et~al.}(2015{\natexlab{a}}){Johnstone}, {G{\"u}del},
  {Brott}, \& {L{\"u}ftinger}}]{Johnstoneetal2015AA577II}
{Johnstone}, C.~P., {G{\"u}del}, M., {Brott}, I., \& {L{\"u}ftinger}, T.
  2015{\natexlab{a}}, \aap, 577, A28

\bibitem[{{Johnstone} {et~al.}(2015{\natexlab{b}}){Johnstone}, {G{\"u}del},
  {L{\"u}ftinger}, {Toth}, \& {Brott}}]{Johnstoneetal2015AA577I}
{Johnstone}, C.~P., {G{\"u}del}, M., {L{\"u}ftinger}, T., {Toth}, G., \&
  {Brott}, I. 2015{\natexlab{b}}, \aap, 577, A27

\bibitem[{{K{\"a}pyl{\"a}} {et~al.}(2014){K{\"a}pyl{\"a}}, {K{\"a}pyl{\"a}}, \&
  {Brandenburg}}]{Kapylaetal2014}
{K{\"a}pyl{\"a}}, P.~J., {K{\"a}pyl{\"a}}, M.~J., \& {Brandenburg}, A. 2014,
  \aap, 570, A43

\bibitem[{Kaula(1964)}]{Kaula1964}
Kaula, W.~M. 1964, Rev. Geophys., 2, 661

\bibitem[{Kawaler(1988)}]{Kawaler1988}
Kawaler, S.~D. 1988, ApJ, 333, 236

\bibitem[{{Lanza} \& {Shkolnik}(2014)}]{LanzaShkolnik2014}
{Lanza}, A.~F. \& {Shkolnik}, E.~L. 2014, \mnras, 443, 1451

\bibitem[{{Laskar} {et~al.}(2012){Laskar}, {Bou{\'e}}, \&
  {Correia}}]{Laskaretal2012}
{Laskar}, J., {Bou{\'e}}, G., \& {Correia}, A.~C.~M. 2012, \aap, 538, A105

\bibitem[{{Leconte} {et~al.}(2010){Leconte}, {Chabrier}, {Baraffe}, \&
  {Levrard}}]{Leconteetal2010}
{Leconte}, J., {Chabrier}, G., {Baraffe}, I., \& {Levrard}, B. 2010, \aap, 516,
  A64

\bibitem[{{Linker} {et~al.}(2017){Linker}, {Caplan}, {Downs}, {Riley}, {Mikic},
  {Lionello}, {Henney}, {Arge}, {Liu}, {Derosa}, {Yeates}, \&
  {Owens}}]{Linkeretal2017}
{Linker}, J.~A., {Caplan}, R.~M., {Downs}, C., {et~al.} 2017, \apj, 848, 70

\bibitem[{MacDonald(1964)}]{MacDonald1964}
MacDonald, G. J.~F. 1964, Rev. Geophys., 2, 467

\bibitem[{MacGregor(1991)}]{MacGregor1991}
MacGregor, K.~B. 1991, in Proceedings of the NATO Advanced Research Workshop on
  Angular Momentum Evolution of Young Stars, 315--331

\bibitem[{MacGregor \& Brenner(1991)}]{MacGregorBrenner1991}
MacGregor, K.~B. \& Brenner, M. 1991, ApJ, 376, 204

\bibitem[{{Mathis}(2013)}]{Mathis2013}
{Mathis}, S. 2013, in Lecture Notes in Physics, Berlin Springer Verlag, Vol.
  865, Lecture Notes in Physics, Berlin Springer Verlag, ed. M.~{Goupil},
  K.~{Belkacem}, C.~{Neiner}, F.~{Ligni{\`e}res}, \& J.~J. {Green}, 23

\bibitem[{Mathis(2015)}]{Mathis2015}
Mathis, S. 2015, A\&A, 580, L3

\bibitem[{{Mathis} {et~al.}(2016){Mathis}, {Auclair-Desrotour}, {Guenel},
  {Gallet}, \& {Le Poncin-Lafitte}}]{Mathisetal2016}
{Mathis}, S., {Auclair-Desrotour}, P., {Guenel}, M., {Gallet}, F., \& {Le
  Poncin-Lafitte}, C. 2016, \aap, 592, A33

\bibitem[{Mathis \& Le~Poncin-Lafitte(2009)}]{MathisLePoncinLafitte2009}
Mathis, S. \& Le~Poncin-Lafitte, C. 2009, A\&A, 497, 889

\bibitem[{{Mathis} \& {Remus}(2013)}]{MathisRemus2013}
{Mathis}, S. \& {Remus}, F. 2013, in Lecture Notes in Physics, Berlin Springer
  Verlag, Vol. 857, Lecture Notes in Physics, Berlin Springer Verlag, ed. J.-P.
  {Rozelot} \& C.~. {Neiner}, 111--147

\bibitem[{Matt {et~al.}(2015)Matt, Brun, Baraffe, Bouvier, \&
  Chabrier}]{Mattetal2015}
Matt, S.~P., Brun, A.-S., Baraffe, I., Bouvier, J., \& Chabrier, G. 2015, ApJL,
  799, L23

\bibitem[{Matt {et~al.}(2012)Matt, Pinzón, Greene, \& Pudritz}]{Mattetal2012a}
Matt, S.~P., Pinzón, G., Greene, T.~P., \& Pudritz, R.~E. 2012, ApJ, 745, 101

\bibitem[{Matt \& Pudritz(2005)}]{MattPudritz2005}
Matt, S.~P. \& Pudritz, R.~E. 2005, ApJ, 632, L135

\bibitem[{Mayor \& Queloz(1995)}]{MayorQueloz1995}
Mayor, M. \& Queloz, D. 1995, Nature, 378, 355

\bibitem[{{McQuillan} {et~al.}(2013{\natexlab{a}}){McQuillan}, {Aigrain}, \&
  {Mazeh}}]{McQuillanAigrainMazeh2013}
{McQuillan}, A., {Aigrain}, S., \& {Mazeh}, T. 2013{\natexlab{a}}, \mnras, 432,
  1203

\bibitem[{{McQuillan} {et~al.}(2013{\natexlab{b}}){McQuillan}, {Mazeh}, \&
  {Aigrain}}]{McQuillanMazehAigrain2013}
{McQuillan}, A., {Mazeh}, T., \& {Aigrain}, S. 2013{\natexlab{b}}, \apjl, 775,
  L11

\bibitem[{{Metcalfe} {et~al.}(2016){Metcalfe}, {Egeland}, \& {van
  Saders}}]{Metcalfeetal2016}
{Metcalfe}, T.~S., {Egeland}, R., \& {van Saders}, J. 2016, \apjl, 826, L2

\bibitem[{Metzger {et~al.}(2012)Metzger, Giannios, \&
  Spiegel}]{Metzgeretal2012}
Metzger, B.~D., Giannios, D., \& Spiegel, D.~S. 2012, MNRAS, 425, 2778

\bibitem[{Mignard(1979)}]{Mignard1979}
Mignard, F. 1979, M\&P, 20, 301

\bibitem[{Murray \& Dermott(1999)}]{MurrayDermott1999}
Murray, C.~D. \& Dermott, S.~F. 1999, Solar System Dynamics (Cambridge
  University Press)

\bibitem[{Noyes {et~al.}(1984)Noyes, Hartmann, Baliunas, Duncan, \&
  Vaughan}]{Noyesetal1984}
Noyes, R.~W., Hartmann, L.~W., Baliunas, S.~L., Duncan, D.~K., \& Vaughan,
  A.~H. 1984, ApJ, 279, 763

\bibitem[{{O'Connor} \& {Hansen}(2018)}]{OconnorHansen2018}
{O'Connor}, C.~E. \& {Hansen}, B.~M.~S. 2018, \mnras, 477, 175

\bibitem[{Ogilvie(2013)}]{Ogilvie2013}
Ogilvie, G.~I. 2013, MNRAS, 429, 613

\bibitem[{Ogilvie(2014)}]{Ogilvie2014}
Ogilvie, G.~I. 2014, ArA\&A, 52, 171

\bibitem[{{Ogilvie} \& {Lesur}(2012)}]{OgilvieLesur2012}
{Ogilvie}, G.~I. \& {Lesur}, G. 2012, \mnras, 422, 1975

\bibitem[{Ogilvie \& Lin(2007)}]{OgilvieLin2007}
Ogilvie, G.~I. \& Lin, D. N.~C. 2007, MNRAS, 661, 1180

\bibitem[{Parker(1958)}]{Parker1958}
Parker, E.~N. 1958, ApJ, 128, 664

\bibitem[{Pizzolato {et~al.}(2003)Pizzolato, Maggio, Micela, Sciortino, \&
  Ventura}]{Pizzolatoetal2003}
Pizzolato, N., Maggio, A., Micela, G., Sciortino, S., \& Ventura, P. 2003,
  A\&A, 397, 147

\bibitem[{{Rasio} {et~al.}(1996){Rasio}, {Tout}, {Lubow}, \&
  {Livio}}]{Rasioetal1996}
{Rasio}, F.~A., {Tout}, C.~A., {Lubow}, S.~H., \& {Livio}, M. 1996, \apj, 470,
  1187

\bibitem[{{Rauer} {et~al.}(2014){Rauer}, {Catala}, {Aerts}, {Appourchaux},
  {Benz}, {Brandeker}, {Christensen-Dalsgaard}, {Deleuil}, {Gizon}, {Goupil},
  {G{\"u}del}, {Janot-Pacheco}, {Mas-Hesse}, {Pagano}, {Piotto}, {Pollacco},
  {Santos}, {Smith}, {Su{\'a}rez}, {Szab{\'o}}, {Udry}, {Adibekyan}, {Alibert},
  {Almenara}, {Amaro-Seoane}, {Eiff}, {Asplund}, {Antonello}, {Barnes},
  {Baudin}, {Belkacem}, {Bergemann}, {Bihain}, {Birch}, {Bonfils}, {Boisse},
  {Bonomo}, {Borsa}, {Brand{\~a}o}, {Brocato}, {Brun}, {Burleigh}, {Burston},
  {Cabrera}, {Cassisi}, {Chaplin}, {Charpinet}, {Chiappini}, {Church},
  {Csizmadia}, {Cunha}, {Damasso}, {Davies}, {Deeg}, {D{\'{\i}}az}, {Dreizler},
  {Dreyer}, {Eggenberger}, {Ehrenreich}, {Eigm{\"u}ller}, {Erikson}, {Farmer},
  {Feltzing}, {de Oliveira Fialho}, {Figueira}, {Forveille}, {Fridlund},
  {Garc{\'{\i}}a}, {Giommi}, {Giuffrida}, {Godolt}, {Gomes da Silva},
  {Granzer}, {Grenfell}, {Grotsch-Noels}, {G{\"u}nther}, {Haswell}, {Hatzes},
  {H{\'e}brard}, {Hekker}, {Helled}, {Heng}, {Jenkins}, {Johansen},
  {Khodachenko}, {Kislyakova}, {Kley}, {Kolb}, {Krivova}, {Kupka}, {Lammer},
  {Lanza}, {Lebreton}, {Magrin}, {Marcos-Arenal}, {Marrese}, {Marques},
  {Martins}, {Mathis}, {Mathur}, {Messina}, {Miglio}, {Montalban}, {Montalto},
  {Monteiro}, {Moradi}, {Moravveji}, {Mordasini}, {Morel}, {Mortier},
  {Nascimbeni}, {Nelson}, {Nielsen}, {Noack}, {Norton}, {Ofir}, {Oshagh},
  {Ouazzani}, {P{\'a}pics}, {Parro}, {Petit}, {Plez}, {Poretti}, {Quirrenbach},
  {Ragazzoni}, {Raimondo}, {Rainer}, {Reese}, {Redmer}, {Reffert},
  {Rojas-Ayala}, {Roxburgh}, {Salmon}, {Santerne}, {Schneider}, {Schou},
  {Schuh}, {Schunker}, {Silva-Valio}, {Silvotti}, {Skillen}, {Snellen}, {Sohl},
  {Sousa}, {Sozzetti}, {Stello}, {Strassmeier}, {{\v S}vanda}, {Szab{\'o}},
  {Tkachenko}, {Valencia}, {Van Grootel}, {Vauclair}, {Ventura}, {Wagner},
  {Walton}, {Weingrill}, {Werner}, {Wheatley}, \& {Zwintz}}]{Raueretal2014}
{Rauer}, H., {Catala}, C., {Aerts}, C., {et~al.} 2014, Experimental Astronomy,
  38, 249

\bibitem[{{Reiners} \& {Schmitt}(2003)}]{ReinersSchmitt2003}
{Reiners}, A. \& {Schmitt}, J.~H.~M.~M. 2003, \aap, 398, 647

\bibitem[{{Reiners} {et~al.}(2014){Reiners}, {Sch{\"u}ssler}, \&
  {Passegger}}]{Reinersetal2014}
{Reiners}, A., {Sch{\"u}ssler}, M., \& {Passegger}, V.~M. 2014, \apj, 794, 144

\bibitem[{Remus {et~al.}(2012)Remus, Mathis, \& Zahn}]{Remusetal2012}
Remus, F., Mathis, S., \& Zahn, J.~P. 2012, A\&A, 544, A132

\bibitem[{{Rudiger} \& {Kitchatinov}(1997)}]{RudigerKitchatinov1997}
{Rudiger}, G. \& {Kitchatinov}, L.~L. 1997, Astronomische Nachrichten, 318, 273

\bibitem[{Réville {et~al.}(2015{\natexlab{a}})Réville, Brun, Matt, Strugarek,
  \& Pinto}]{Revilleetal2015a}
Réville, V., Brun, A.-S., Matt, S.~P., Strugarek, A., \& Pinto, R.~F.
  2015{\natexlab{a}}, ApJ, 798, 116

\bibitem[{Réville {et~al.}(2015{\natexlab{b}})Réville, Brun, Strugarek, Matt,
  Bouvier, Folsom, \& Petit}]{Revilleetal2015b}
Réville, V., Brun, A.-S., Strugarek, A., {et~al.} 2015{\natexlab{b}}, ApJ,
  814, 99

\bibitem[{Réville {et~al.}(2016)Réville, Folsom, Strugarek, \&
  Brun}]{Revilleetal2016}
Réville, V., Folsom, C.~P., Strugarek, A., \& Brun, A.-S. 2016, ApJ, 832, 145

\bibitem[{{Sadeghi Ardestani} {et~al.}(2017){Sadeghi Ardestani}, {Guillot}, \&
  {Morel}}]{SadeghiArdestanietal2017}
{Sadeghi Ardestani}, L., {Guillot}, T., \& {Morel}, P. 2017, \mnras, 472, 2590

\bibitem[{Schatten {et~al.}(1969)Schatten, Wilcox, \& Ness}]{Schattenetal1969}
Schatten, K.~H., Wilcox, J.~M., \& Ness, N.~F. 1969, SoPh, 6, 442

\bibitem[{Schatzman(1962)}]{Schatzman1962}
Schatzman, E. 1962, AnAp, 25, 18

\bibitem[{{Schrijver} \& {DeRosa}(2003)}]{SchrijverDeRosa2003}
{Schrijver}, C.~J. \& {DeRosa}, M.~L. 2003, \solphys, 212, 165

\bibitem[{{See} {et~al.}(2017){See}, {Jardine}, {Vidotto}, {Donati}, {Boro
  Saikia}, {Fares}, {Folsom}, {H{\'e}brard}, {Jeffers}, {Marsden}, {Morin},
  {Petit}, {Waite}, \& {BCool Collaboration}}]{Seeetal2017}
{See}, V., {Jardine}, M., {Vidotto}, A.~A., {et~al.} 2017, \mnras, 466, 1542

\bibitem[{Skumanich(1972)}]{Skumanich1972}
Skumanich, A. 1972, ApJ, 171, 565

\bibitem[{{Spada} {et~al.}(2013){Spada}, {Demarque}, {Kim}, \&
  {Sills}}]{Spadaetal2013}
{Spada}, F., {Demarque}, P., {Kim}, Y.-C., \& {Sills}, A. 2013, \apj, 776, 87

\bibitem[{{Spada} {et~al.}(2010){Spada}, {Lanzafame}, \&
  {Lanza}}]{Spadaetal2010}
{Spada}, F., {Lanzafame}, A.~C., \& {Lanza}, A.~F. 2010, \mnras, 404, 641

\bibitem[{{Spiegel} \& {Zahn}(1992)}]{SpiegelZahn1992}
{Spiegel}, E.~A. \& {Zahn}, J.-P. 1992, \aap, 265, 106

\bibitem[{{Strugarek } {et~al.}(2011){Strugarek }, {Brun}, \&
  {Zahn}}]{Strugareketal2011b}
{Strugarek }, A., {Brun}, A.~S., \& {Zahn}, J.-P. 2011, Astronomische
  Nachrichten, 332, 891

\bibitem[{{Strugarek} {et~al.}(2017{\natexlab{a}}){Strugarek}, {Beaudoin},
  {Charbonneau}, {Brun}, \& {do Nascimento}}]{Strugareketal2017Sci}
{Strugarek}, A., {Beaudoin}, P., {Charbonneau}, P., {Brun}, A.~S., \& {do
  Nascimento}, J.-D. 2017{\natexlab{a}}, Science, 357, 185

\bibitem[{{Strugarek} {et~al.}(2017{\natexlab{b}}){Strugarek}, {Bolmont},
  {Mathis}, {Brun}, {R{\'e}ville}, {Gallet}, \&
  {Charbonnel}}]{Strugareketal2017}
{Strugarek}, A., {Bolmont}, E., {Mathis}, S., {et~al.} 2017{\natexlab{b}},
  \apjl, 847, L16

\bibitem[{Strugarek {et~al.}(2014)Strugarek, Brun, Matt, \&
  Réville}]{Strugareketal2014}
Strugarek, A., Brun, A.~S., Matt, S.~P., \& Réville, V. 2014, ApJ, 795, 86 (17
  pp.)

\bibitem[{Strugarek {et~al.}(2015)Strugarek, Brun, Matt, \&
  Réville}]{Strugareketal2015}
Strugarek, A., Brun, A.~S., Matt, S.~P., \& Réville, V. 2015, ApJ, 815, 111
  (14 pp.)

\bibitem[{{Suzuki} {et~al.}(2013){Suzuki}, {Imada}, {Kataoka}, {Kato},
  {Matsumoto}, {Miyahara}, \& {Tsuneta}}]{Suzukietal2013}
{Suzuki}, T.~K., {Imada}, S., {Kataoka}, R., {et~al.} 2013, \pasj, 65, 98

\bibitem[{{Talon} \& {Charbonnel}(2005)}]{TalonCharbonnel2005}
{Talon}, S. \& {Charbonnel}, C. 2005, \aap, 440, 981

\bibitem[{{Teitler} \& {K{\"o}nigl}(2014)}]{TeitlerKonigl2014}
{Teitler}, S. \& {K{\"o}nigl}, A. 2014, \apj, 786, 139

\bibitem[{{Terquem} {et~al.}(1998){Terquem}, {Papaloizou}, {Nelson}, \&
  {Lin}}]{Terquemetal1998}
{Terquem}, C., {Papaloizou}, J.~C.~B., {Nelson}, R.~P., \& {Lin}, D.~N.~C.
  1998, \apj, 502, 788

\bibitem[{{Trilling} {et~al.}(1998){Trilling}, {Benz}, {Guillot}, {Lunine},
  {Hubbard}, \& {Burrows}}]{Trillingetal1998}
{Trilling}, D.~E., {Benz}, W., {Guillot}, T., {et~al.} 1998, \apj, 500, 428

\bibitem[{{van Saders} {et~al.}(2016){van Saders}, {Ceillier}, {Metcalfe},
  {Silva Aguirre}, {Pinsonneault}, {Garc{\'{\i}}a}, {Mathur}, \&
  {Davies}}]{vanSadersetal2016}
{van Saders}, J.~L., {Ceillier}, T., {Metcalfe}, T.~S., {et~al.} 2016, \nat,
  529, 181

\bibitem[{Vidotto {et~al.}(2014)Vidotto, Gregory, Jardine, Donati, Morin,
  Folsom, Bouvier, Cameron, Hussain, Marsden, Waite, Fares, Jeffers, \&
  do~Nascimento}]{Vidottoetal2014}
Vidotto, A.~A., Gregory, S.~G., Jardine, M., {et~al.} 2014, MNRAS, 441, 2361

\bibitem[{Weber \& Davis(1967)}]{WeberDavis1967}
Weber, E.~J. \& Davis, Jr., L. 1967, ApJ, 148, 217

\bibitem[{{Witte} \& {Savonije}(2002)}]{WitteSavonije2002}
{Witte}, M.~G. \& {Savonije}, G.~J. 2002, \aap, 386, 222

\bibitem[{{Wood} {et~al.}(2018){Wood}, {Laming}, {Warren}, \&
  {Poppenhaeger}}]{Woodetal2018}
{Wood}, B.~E., {Laming}, J.~M., {Warren}, H.~P., \& {Poppenhaeger}, K. 2018,
  \apj, 862, 66

\bibitem[{Wright {et~al.}(2011)Wright, Drake, Mamajek, \&
  Henry}]{Wrightetal2011}
Wright, N.~J., Drake, J.~J., Mamajek, E.~E., \& Henry, G.~W. 2011, ApJ, 743, 48

\bibitem[{Zahn(1966)}]{Zahn1966b}
Zahn, J.~P. 1966, AnAp, 29, 489

\bibitem[{Zahn(1975)}]{Zahn1975}
Zahn, J.~P. 1975, A\&A, 41, 329

\bibitem[{Zahn(1977)}]{Zahn1977}
Zahn, J.~P. 1977, A\&A, 57, 383

\bibitem[{Zahn(1989)}]{Zahn1989}
Zahn, J.~P. 1989, A\&A, 220, 112

\bibitem[{{Zahn}(1992)}]{Zahn1992}
{Zahn}, J.-P. 1992, \aap, 265, 115

\bibitem[{Zanni \& Ferreira(2013)}]{ZanniFerreira2013}
Zanni, C. \& Ferreira, J. 2013, A\&A, 550, A99

\bibitem[{Zhang \& Penev(2014)}]{ZhangPenev2014}
Zhang, M. \& Penev, K. 2014, ApJ, 787, 131 (7 pp.)

\end{thebibliography}

\appendix

\section{Description of the code}
\label{sec:descriptionCode}

\begin{figure*}
\centering
\includegraphics[width=17cm]{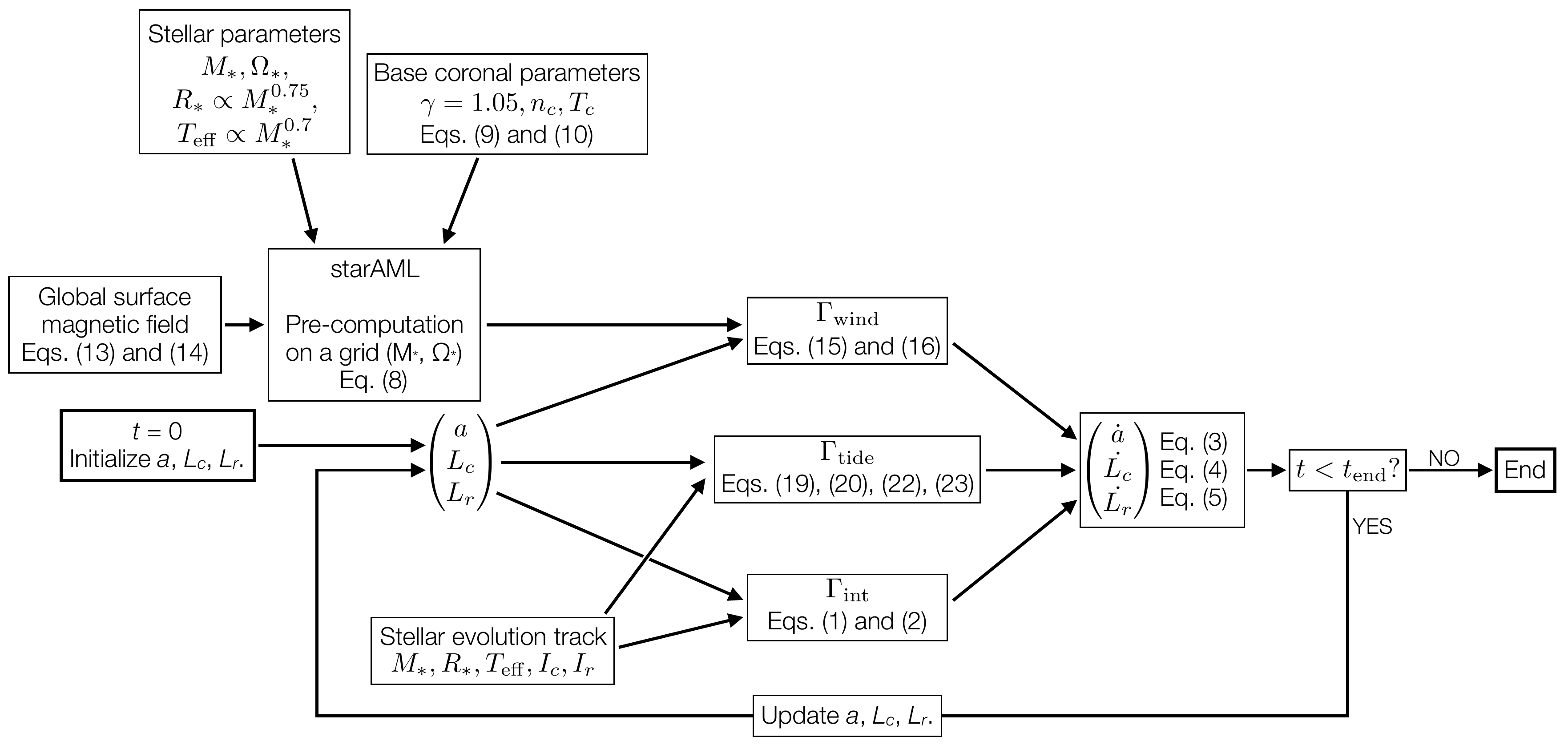}
\caption{Scheme of the dependencies in the code. We first initialize the values of $a$, $L_c$, and $L_r$. At each time step, the torques are computed from the current state of the system, the stellar evolution variables, and the results of starAML. Then, the time derivatives $\dot{a}$, $\dot{L}_c$, and $\dot{L}_r$ are computed from the torques. This iterative integration goes on until the age of the system is greater than the limit age we set for the computation.}
\label{fig:appendixScheme}
\end{figure*}

This appendix summarizes the key equations in the paper that are solved by the code. The purpose of ESPEM is to compute three quantities of interest over the life of the considered star-planet system: the semi-major axis $a$, the angular momentum of the shellular convective envelope $L_c$, and that of the inner radiative zone $L_r$. At each time step, the torques $\Gint$, $\Gwind$, and $\Gtide$ are computed from the system's state. The time derivatives of the quantities of interest are then computed from these torques.

The computation of $\Gwind$ is different from that of the other two torques. To compute the stellar wind, we used the starAML code which implements the method of \citet{Revilleetal2015b}. We refer the reader to section~\ref{sec:rotation} from more details about this method. Since their technique implies to compute the radial profile of the wind and magnetic field in the whole stellar corona, we decided to precompute the torque of the wind rather than calling the starAML routine at each ESPEM time step because the latter option would have significantly slowed down our simulations.

The other two torques, $\Gint$ and $\Gtide$, are directly computed from the system's state and the stellar evolution variables. The latter are read from a precomputed STAREVOL evolutionary track \citep{Amardetal2016}.

\section{Comparison to other models}
\label{sec:compareOldModels}

This work is similar to other studies of star-planet systems. In this appendix, we discuss the differences brought by the tidal prescription and the two-layer internal rotation model we implemented in ESPEM. To that end, we compare the results of these frameworks with those of state-of-the-art models.

\subsection{Comparison with the constant quality factor model}
\label{subsec:compare_ZP}

In this paragraph, we emphasize the consequences of changing the tidal prescription. We compare the predictions made by the model described in section~\ref{sec:tides} with those of the constant quality factor model, used for instance by \citet{ZhangPenev2014}. For this purpose, we computed for each tidal prescription, the evolution of a system composed by a solar-mass star and a Jupiter-mass planet with an initial semi-major axis equal to 0.03 AU. The initial rotation period of the star was set to 1.4 days, which corresponds to the fast rotator described by \citet{GalletBouvier2015}. We tested three values of $Q'$ for the constant quality factor prescription, $10^6$, $10^7$ and $10^8$. Here, we detail the case where $Q' = 10^6$. The two other scenarios are discussed at the end of this subsection.

\begin{figure} 
\centering
\includegraphics[width=\hsize]{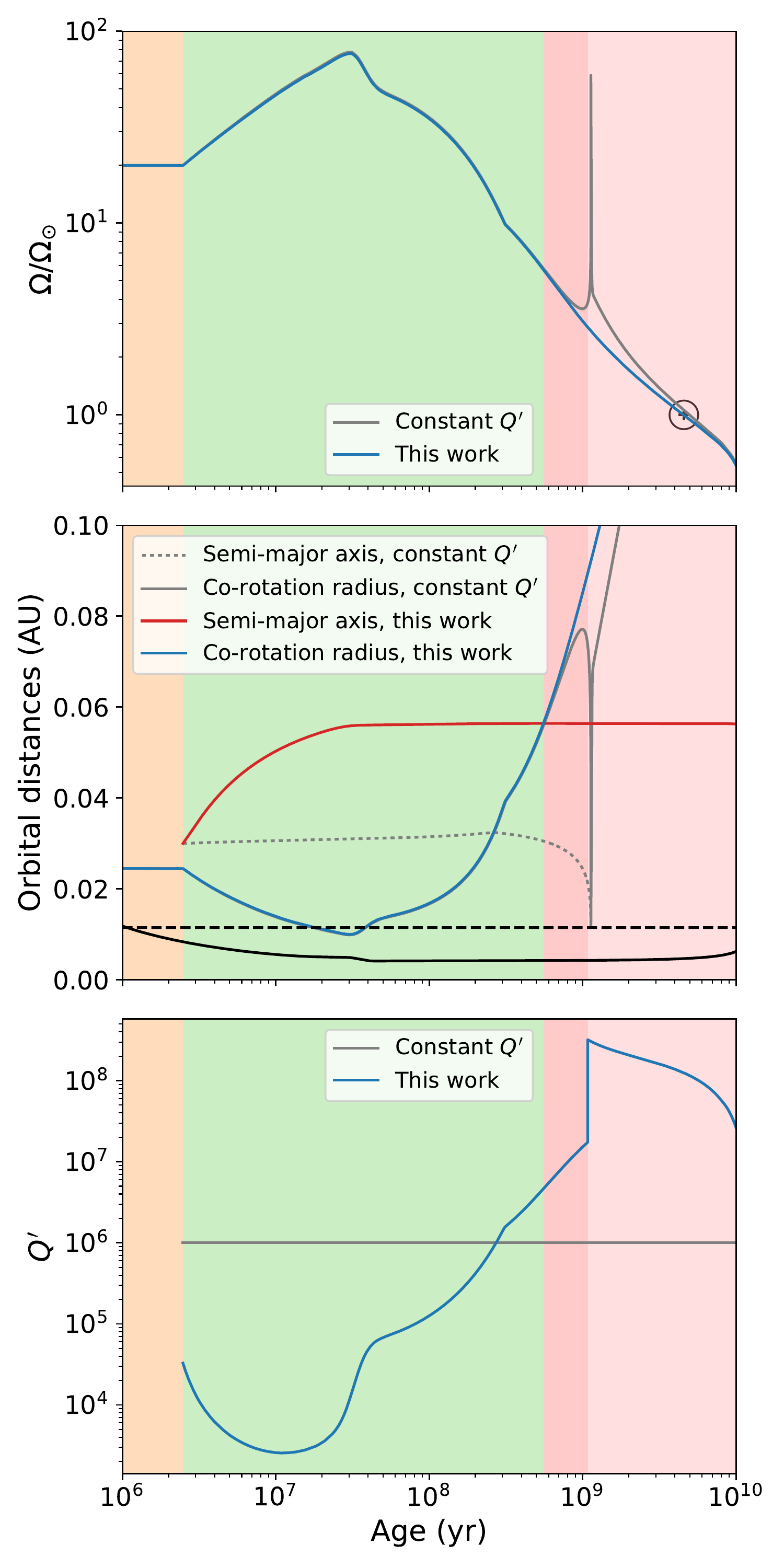}
\caption{Evolution of a system constituted by a star of mass 1 $\Msun$ and a planet of mass 1 $\Mjup$ in the constant quality factor model with $Q' = 10^6$. We chose an initial semi-major axis $\aini$ = 0.03 AU and an initial stellar rotation period $\Pini$ = 1.4 days. Color background mark the evolution phases in the varaible quality factor case. \textit{Orange background:} disk-locking phase. \textit{Green background:} outward planet migration driven by tidal inertial waves. \textit{Pink background:} inward planet migration under the combined equilibrium and dynamical tides. \textit{Light pink background:} inward planet migration under the equilibrium tide.}
\label{fig:compare_ZP_appx}
\end{figure}

Fig.~\ref{fig:compare_ZP_appx} shows the comparison between the constant $Q'$ and variable $Q'$ models. As can be seen in the middle panel, the former case leads to the destruction of the planet whereas in the latter the planet survives. What allows the survival in the variable $Q'$ model is the fast outward migration during the first million years of the simulation. This phenomenon is due to the low value of the tidal quality factor in this period.

The color backgrounds correspond to the different evolution phases of the system in the variable $Q'$ case. The orange background stands for the disk-locking phase. The green background represents the outward-migration phase. During this stage, both the equilibrium and the dynamical tides are active. The pink background corresponds to the inward-migration phase during which both tides are raised. During this stage, the orbital period $\Porb$ is such that $P_* / 2 < \Porb < P_*$, where $P_*$ is the surface rotation of the star. Finally, the light pink background depicts the inward-migration phase during which only the equilibrium tide is active. During the latter two phases, the planet is too far from the star to undergo significant orbital migration.

This figure illustrates the difference between our tidal model and the constant $Q'$. \citet{BolmontMathis2016} performed simulations on a extensive parameter range to show the consistency of this difference.

\subsection{Comparison with the one-layer rotation model}
\label{subsec:compare_BM}

One of the main differences between our code and that of \citet{BolmontMathis2016} is the internal rotation model implemented. We used the two-layer model of \citet{MacGregorBrenner1991}, in which the tidal and wind braking torques are applied on the convective envelope, which relays them to the core through a parametrized internal coupling torque with a constant timescale, whereas they assume that the star is in solid body rotation. Consequently, stellar rotation in our model is expected to be more sensible to tidal dissipation due to the low inertia of the convective envelope. To investigate this, we measured the spin-down of a star caused by a close-in massive planet in orbit beyond the co-rotation radius. For both the one-layer and the two-layer models, we computed the secular evolution of the rotation period of a 0.6 $\Msun$ star such as $\Pini$ = 1.2 days, in a first time without planet, then with a 5 $\Mjup$ planet initially orbiting at a distance $\aini$ = 0.024 AU. We then calculated the difference $\delta P$ between the two cases as defined in Eq.~\eqref{eq:defDeltaP}.

\begin{figure}
\centering
\includegraphics[width=\hsize]{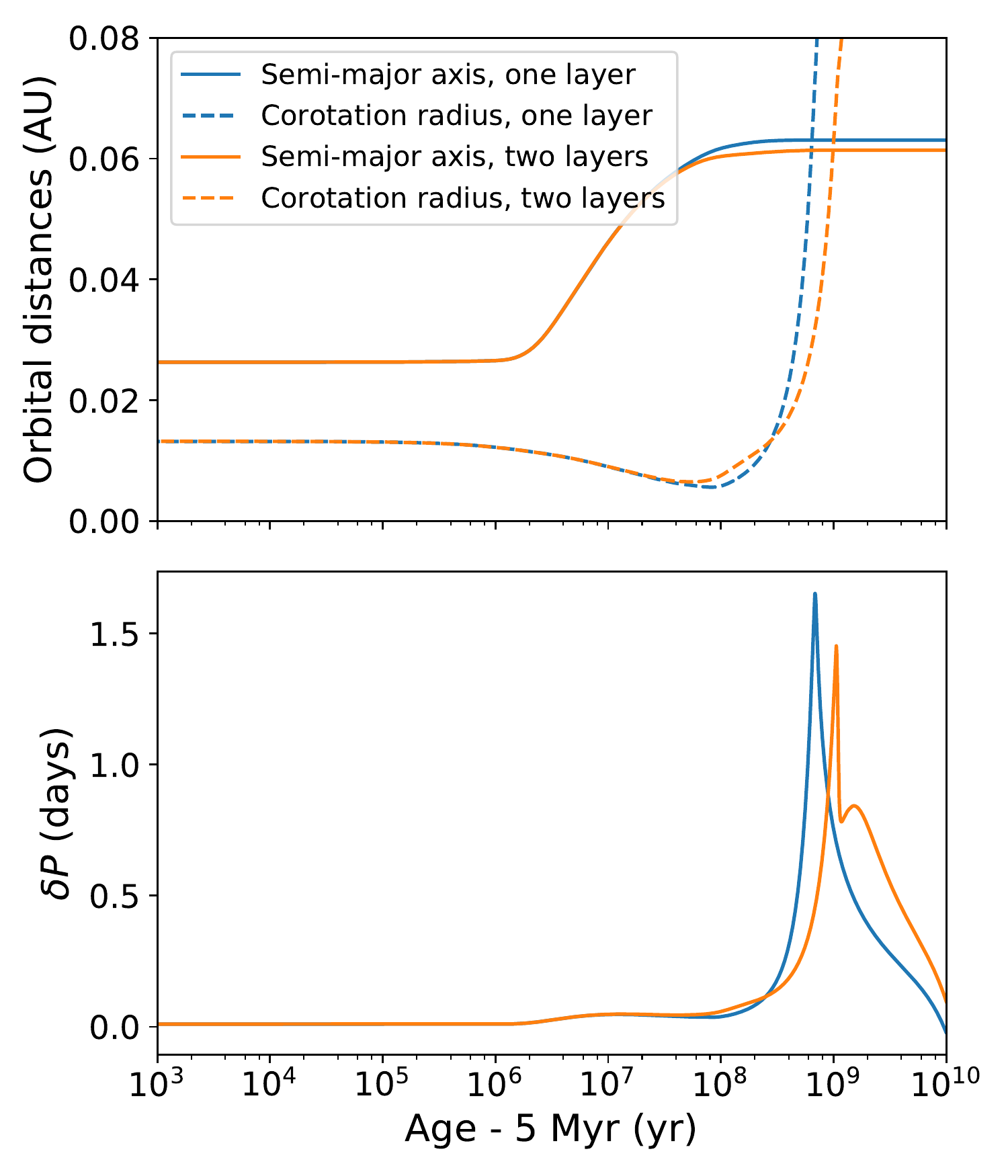}
\caption{Comparison between the one-layer (blue) and the two-layer (orange) models. Secular evolution of a system formed by a 0.6 $\Msun$ star with initial rotation period $\Pini$ = 1.2 days and a 5 $\Mjup$ planet initially orbiting at a distance $\aini$ = 0.024 AU. \textit{Top panel:} semi-major axis (solid lines) and co-rotation radius of the star (dashed lines). \textit{Bottom panel:} Difference to the rotation period in the case without the planet, as defined in Eq.~\eqref{eq:defDeltaP}.}
\label{fig:compare_BM_secular}
\end{figure}

Fig.~\ref{fig:compare_BM_secular} shows that both models predict similar evolutions. As can be seen in the top panel, the planet migrates outward during the first hundreds of millions of years. Consequently, the star is spun down, which is why $\delta P$ is positive. In both cases, $\delta P$ increases during the outward-migration phase, reaches a maximum of the order of one day around the age 1 Gyr, and decreases over the main-sequence phase. The curve of $\delta P$ in the two-layer model is slightly delayed compared to that of the one-layer model. Studying this delay requires to perform simulations on an extensive parameter space, which is beyond the scope of this work.

\section{Discontinuities of the tidal torque}

The formulation of the tidal torque in Eq.~\eqref{eq:tidal_torque} implies discontinuities. This occurs at two events: when the planet orbits the star at a distance close to the co-rotation radius and when the orbital period is close to half the rotation period. In the former case, the tidal frequency is close to zero and the sign of the tidal torque may change with the absolute value remaining constant. In the latter case, the tidal frequency approaches $-2\ \Omstar$, the limit of application of the dynamical tide, where the torque may discontinuously vary by several orders of magnitude, to reach the values of the equilibrium tide.

\begin{figure}
\centering
\includegraphics[width=\hsize]{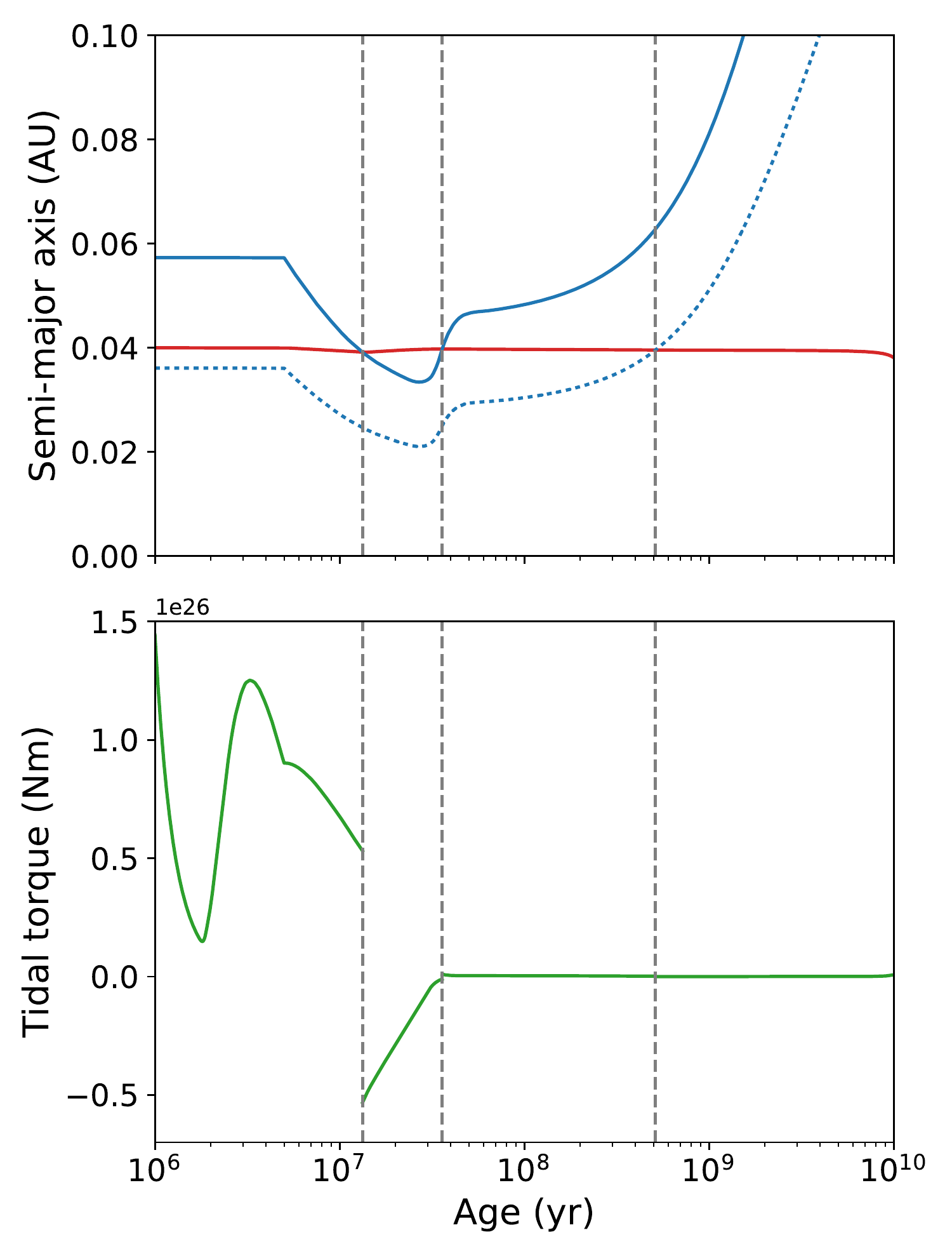}
\caption{Secular evolution of a star-planet system formed by a star of mass 1 $M_{\odot}$ and a planet of mass 1 $M_{\mathrm{Jup}}$. Initial semi-major axis 0.04 AU and initial stellar rotation period 5 days. \textit{Top panel:} Semi-major axis (red line), co-rotation radius (solid blue line) and limit of application of the dynamical tide (dotted blue line). The vertical grey lines correspond to events at which occur discontinuities of the tidal torque. \textit{Bottom panel:} tidal torque.}
\label{fig:Q_secular}
\end{figure}

Figure~\ref{fig:Q_secular} illustrates the different phases of the secular evolution of a star planet-system formed by a star of mass 1 $\Msun$ and a planet of mass 1 $\Mjup$ initially orbiting its host at 0.04 AU. The events at which the discontinuities of the tidal torque occur are marked by vertical dashed grey lines. As is visible on the bottom panel, changes of tidal regime induce strong variations of the tidal torque.

These discontinuities generate numerical problems during the integration of differential equations. The Bulirsch-Stoer algorithm from \citet{Haireretal2000} implements an adaptive step size to ensure the stability of the solution. Discontinuities in the derivatives make the step size shrink and prevent the resolution over the evolution. This is why we slightly modified the dependence of the tidal torque on $\omtide$ in the code.

\begin{figure}
\centering
\includegraphics[width=\hsize]{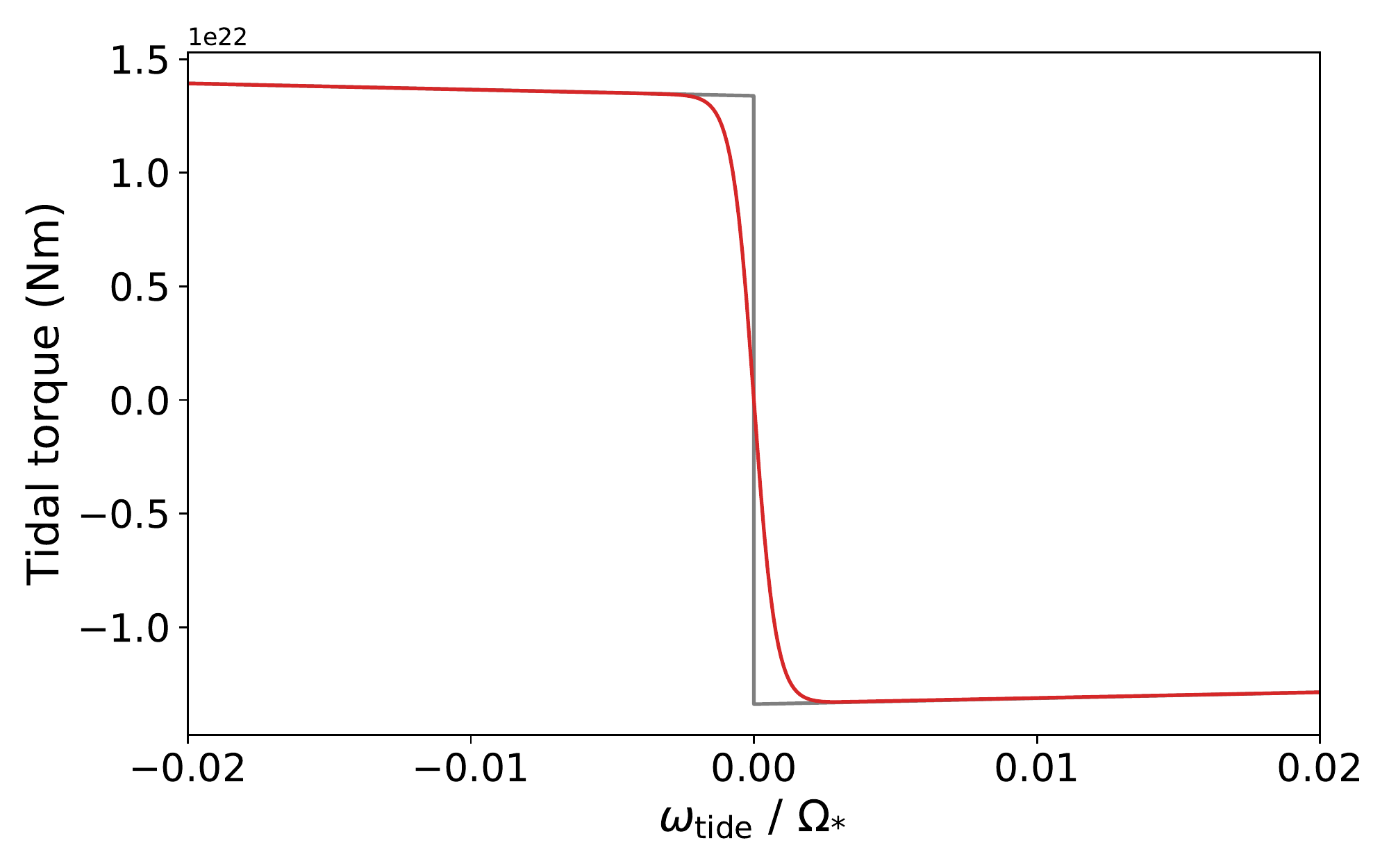}
\caption{Tidal torque around $\omega_{\mathrm{tide}} = 0$ for a system formed by a star of mass 1 $\Msun$ rotating at 5 $\Omsun$ and a planet of mass $\Mjup$. The grey curve represents the theoretical tidal torque and the red curve corresponds to our regularized expression.}
\label{fig:tidal_torque_zero_frequency}
\end{figure}

To fix the null-tidal-frequency situation, we replaced the sign function by a hyperbolic tangent in Eq.~\eqref{eq:tidal_torque}. We introduced the characteristic pulsation $\omega_0 = 10^{-8}$~rad.s$^{-1}$ such that the expression used in ESPEM is:
\begin{equation}
\Gamma_{\mathrm{tide}} = - \mathrm{tanh} \left(\frac{\omtide}{\omega_0}\right) \frac{9}{4 Q'} \frac{G m_{\mathrm{p}}^2}{a^6} R_*^5.
\end{equation}
This modification makes the torque change sign continuously without altering its value for $|\omtide| > |\omega_0|$. For example, Fig.~\ref{fig:tidal_torque_zero_frequency} shows the tidal torque as a function of the tidal frequency around 0 for a system formed by a star of mass 1~$\Msun$ and a planet of mass 1~$\Mjup$.

We performed a similar correction in the second case. For $\omtide < - 2\ \Omstar$, the dynamical tide does not apply so only the equilibrium tide contributes to the orbital evolution. For $\omtide > -2 \ \Omstar$, inertial waves are raised in the stellar envelope and both tides contribute to the interaction. Consequently, the theoretical dissipation can be written as\footnote{Here, we do not have a condition around $\omtide = 2\ \Omega_*$ because this corresponds to a null mean motion, which is never realized under our hypotheses.}:
\begin{align}
\frac{1}{Q'_{\mathrm{th}}} &= \frac{1}{Q'_{\mathrm{eq}}} & \ \mathrm{if}\ \omtide < - 2 \ \Omstar \\
\frac{1}{Q'_{\mathrm{th}}} &= \frac{1}{Q'_{\mathrm{eq}}} + \frac{1}{Q'_{\mathrm{dyn}}} & \ \mathrm{if}\ \omtide > - 2 \ \Omstar.
\end{align}
The discontinuity arises from the addition of the dissipation of the dynamical tide $1 / \Qdyn$ as soon as the tidal frequency passes the limit $2 \ \Omstar$. To fix this problem, we replace this dependence with a more regular function based on a hyperbolic tangent:
\begin{equation}
\frac{1}{Q'_{\mathrm{reg}}} = \frac{1}{Q'_{\mathrm{eq}}} + \frac{1}{Q'_{\mathrm{dyn}}} \frac{1}{2} \left(1 + \tanh\left(\frac{\omtide+ 2 \ \Omstar}{\omega_1}\right)\right)
\end{equation}
where $Q'_{\mathrm{reg}}$ is the regularized tidal quality factor and $\omega_1 = 10^{-10}$~rad.s$^{-1}$. The factor multiplying $1 / \Qdyn$ is null when the inertial waves are not raised and it equals 1 when they apply. Consequently, this regularized dissipation is continuous and fits the theoretical dissipation when the tidal frequency is farther than $\omega_1$ from $-2\ \Omstar$.

The choice of $\omega_1$ is worth commenting upon. To this end, we computed the secular evolution of given star-planet systems for different values of $\omega_1$ ranging from $10^{-10}$ to $10^{-6}$~rad.s$^{-1}$. We did not try larger values for $\omega_1$ because this parameter has to be small compared to $\Omstar$, which is of the order of $\Omsun$ on the main sequence for a solar-like star. Here, we illustrate the results of this experiment with the cases $\omega_1 = 10^{-10}$~rad.s$^{-1}$ and $\omega_1 = 10^{-6}$~rad.s$^{-1}$ for a 1 $\Msun$ star and a 1 $\Mjup$ planet with $\Pini$ = 5 days and $\aini$ = 0.025 AU.

\begin{figure}
\centering
\includegraphics[width=\hsize]{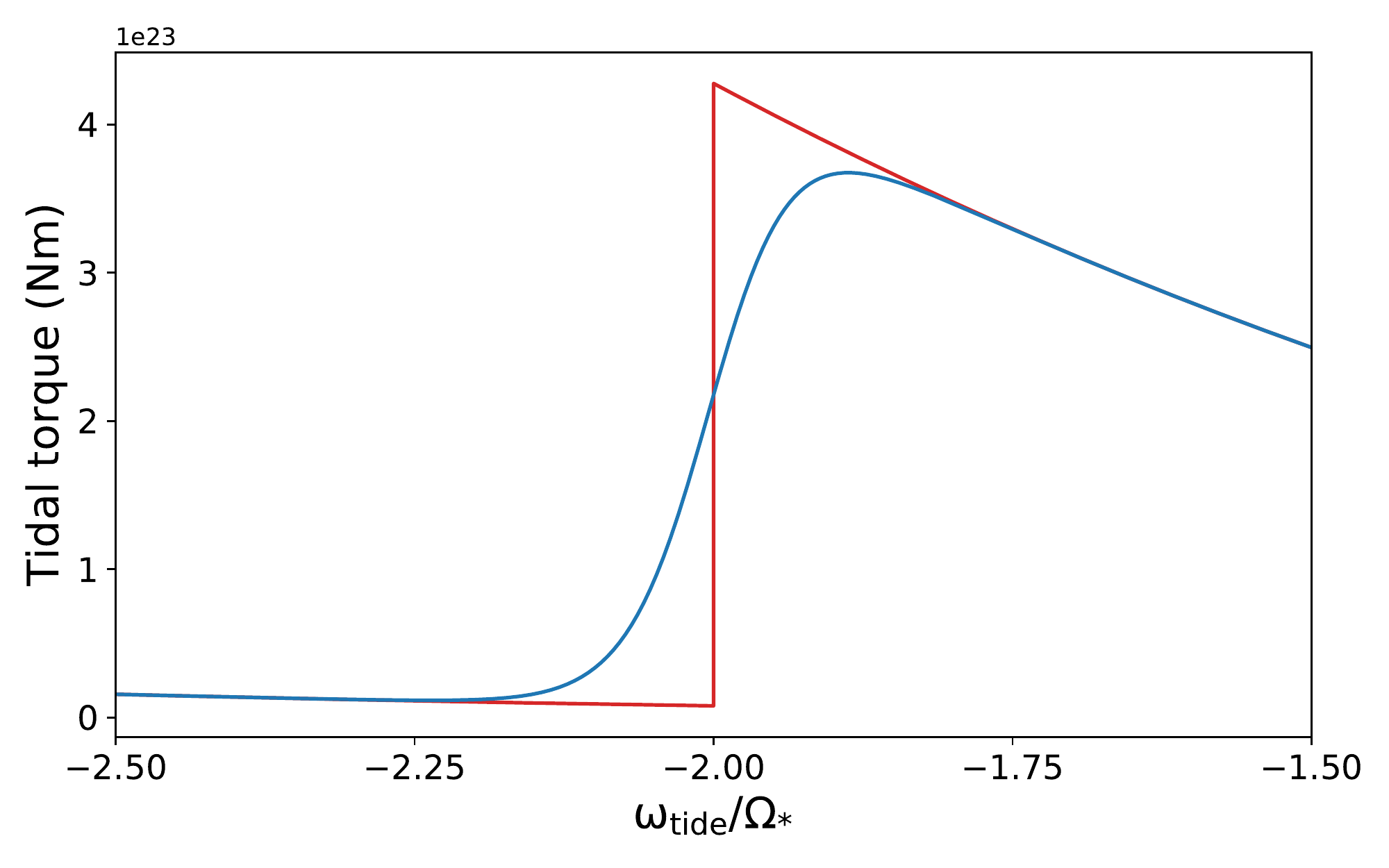}
\caption{Tidal torque around $\omega_{\mathrm{tide}} = -2\ \Omstar$ for a system formed by a star of mass 1 $\Msun$ rotating at 5 $\Omsun$ and a planet of mass $\Mjup$. The red curve represents the tidal torque regularized with $\omega_1 = 10^{-10}$~rad.s$^{-1}$ and the blue curve to $\omega_1 = 10^{-6}$~rad.s$^{-1}$.}
\label{fig:th_test_tidal_limit}
\end{figure}

\begin{figure}
\centering
\includegraphics[width=\hsize]{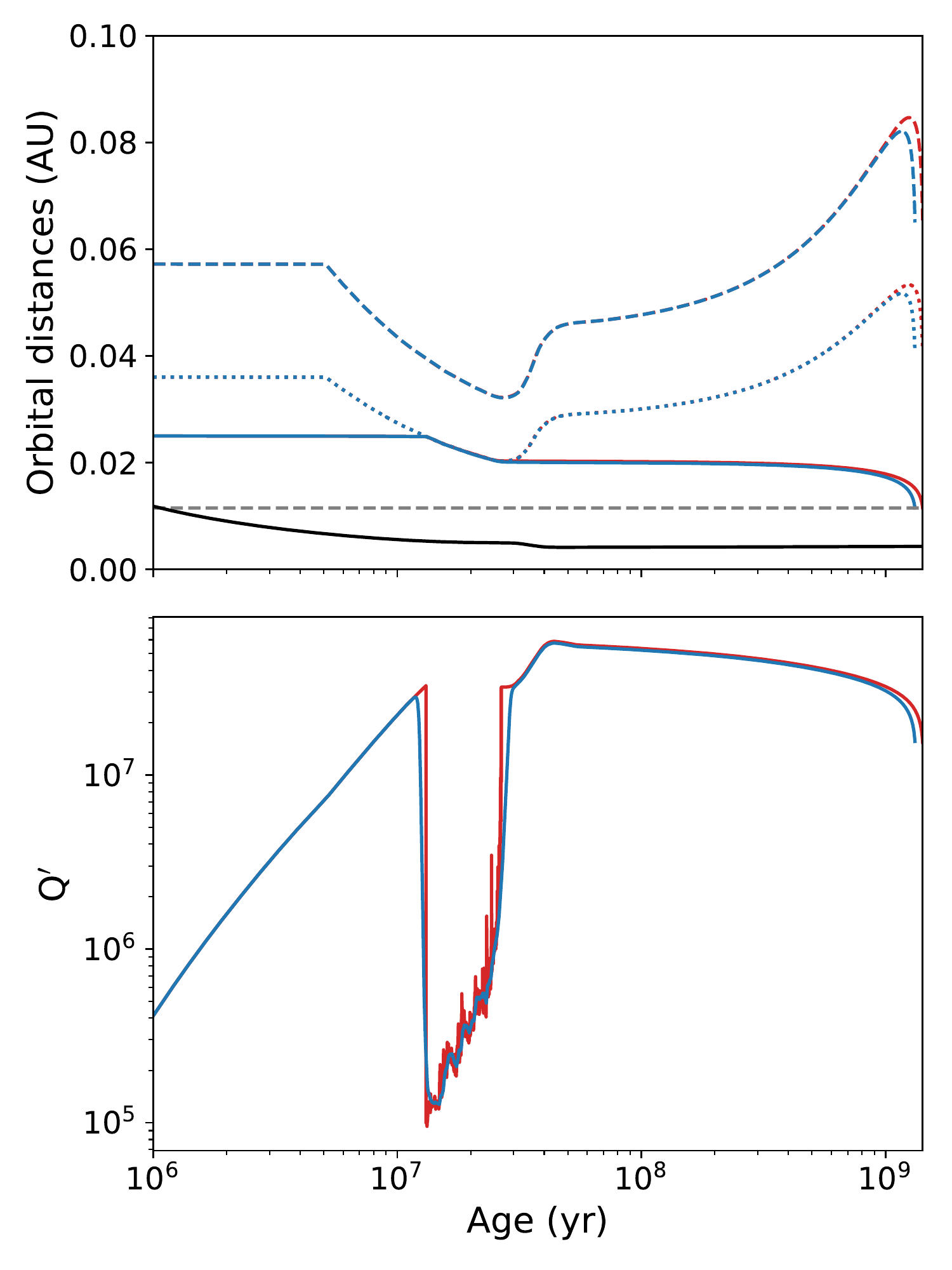}
\caption{Secular evolution of a star-planet system composed by a 1 $\Msun$ star and a 1 $\Mjup$ planet with an initial stellar rotation period $\Pini$ = 5 days and an initial semi-major axis $\aini$ = 0.025 AU. Red curves were computed for $\omega_1 = 10^{-10}$ rad/s and blue curves, for $\omega_1 = 10^{-6}$ rad/s. \textit{Top panel:} Semi-major axis (solid lines), stellar co-rotation radius (dashed lines), limit of excitation of tidal inertial waves (dotted lines), stellar radius (solid black line) and Roche limit (dashed grey line). \textit{Bottom panel:} Modified tidal quality factor.}
\label{fig:th_test_evol}
\end{figure}

Fig.~\ref{fig:th_test_tidal_limit} shows the tidal torque in each case as a function of tidal frequency. Despite the fact that the torque is significantly smoothed for the larger value of $\omega_1$, the evolution of the system is only weakly influenced by this parameter. This can be seen in Fig.~\ref{fig:th_test_evol}. Indeed, in both cases, the evolution phases of the system are the same. The planet's lifetime is slightly shorter in the smoothed case but it is not significantly different from the other. We conclude from this study that the parameter $\omega_1$ has a weak influence on the evolution. Therefore, we set its value to $\omega_1 = 10^{-10} \ \mathrm{rad.s}^{-1}$, so as to stay close to the theoretical tidal calculations. However, the excitation and disappearance of the dynamical tide should be discussed in a further work.

\end{document}